\documentclass[prd,superscriptaddress,nofootinbib,preprintnumbers,amsmath,amssymb,10pt,twocolumn,dvipsnames,floatfix]{revtex4-2}
\usepackage[utf8]{inputenc}
\usepackage{amsfonts}
\usepackage{mathrsfs}
\usepackage{physics}
\usepackage{footnote}
\usepackage{scrextend}
\usepackage{leftidx}
\usepackage{soul}
\usepackage{braket}
\usepackage{makecell}
\usepackage{multirow}
\usepackage{array}
\usepackage{orcidlink} 
\usepackage{bm}
\usepackage[lofdepth,lotdepth,caption=false,justification=RaggedRight]{subfig}
\usepackage{slashed}
\usepackage[capitalize]{cleveref}

\usepackage{xcolor}
\definecolor{red}{rgb}{0.9, 0,0}
\definecolor{cerulean}{rgb}{0., 0.42,0.9}

\usepackage{epsfig}
\usepackage{graphicx}  
\usepackage{url}
\usepackage{color}
\usepackage{hyperref}
\hypersetup{
     colorlinks   = true,
     citecolor    = red,
	 linkcolor=blue
}
\usepackage{float}
\usepackage[export]{adjustbox}
\usepackage{mathrsfs}

\usepackage[ruled,vlined]{algorithm2e}
\usepackage{tikz}
\usetikzlibrary{math}
\usetikzlibrary{decorations.pathreplacing}
\usetikzlibrary{quantikz}

\usepackage[capitalize]{cleveref}    
  \crefname{algorithm}{Alg.}{Algs.}
  \crefname{tab}{Table}{Tables}
  \crefname{fig}{Fig.}{Figs.}
  \crefname{section}{Sec.}{Secs.}
  \crefname{appendix}{App.}{Apps.}
     
\allowdisplaybreaks

\newcolumntype{C}[1]{>{\centering\arraybackslash}p{#1}}
\graphicspath{{./figures/}}

\def\over#1{#1^{\mathrm{over}}}
\newcommand{\labels}{{(q,\bm{c})}}
\newcommand{\pars}{{(q,\bm{c};z)}}
\newcommand{\cpars}{{(\bm{c};z)}}
\newcommand{\intpars}{{(q, \bm{c})}}
\newcommand{\intcpars}{{(\bm{c})}}
\newcommand{\bmc}{\bm{c}}
\newcommand{\bmq}{\bm{q}}
\newcommand{\bmr}{\bm{r}}
\newcommand{\bmh}{\bm{h}}

\newcommand{\qSq}{{\tilde{q}^2}}

\newcommand{\qSqj}{\tilde{q}_j^2}

\newcommand{\Pover}[1][]{{P_{#1}^{\mathrm{over}}}}
\newcommand{\zmover}{{z_-^{\mathrm{over}}}}
\newcommand{\zpover}{{z_+^{\mathrm{over}}}}
\newcommand{\alpover}[1][]{{\alpha_{#1}^{\mathrm{over}}}}

\newcommand{\Ndisc}{{N_z}}
\newcommand{\stateLabel}{s}
\newcommand{\stateReg}[1][]{{\ket{\stateLabel_{#1}}}}

\newboolean{compileAll}
\setboolean{compileAll}{true} 
 
\usepackage{ulem}

\begin{document}


\title{Quantum Parton Shower with Kinematics}

\author{Christian W. Bauer\,\orcidlink{0000-0001-9820-5810}}
\affiliation{Theoretical Physics Group, Lawrence Berkeley National Laboratory, Berkeley, CA 94720, USA}
\affiliation{Berkeley Center for Theoretical Physics, Department of Physics, University of California, Berkeley, CA 94720, USA}
\author{So Chigusa\,\orcidlink{0000-0001-6005-4447}}
\affiliation{Theoretical Physics Group, Lawrence Berkeley National Laboratory, Berkeley, CA 94720, USA}
\affiliation{Berkeley Center for Theoretical Physics, Department of Physics, University of California, Berkeley, CA 94720, USA}
\author{Masahito Yamazaki\,\orcidlink{0000-0001-7164-8187}}
\affiliation{Kavli IPMU (WPI), UTIAS, The University of Tokyo, Kashiwa, Chiba 277-8583, Japan}
\affiliation{Center for Data-Driven Discovery, Kavli IPMU (WPI), UTIAS, The University of Tokyo, Kashiwa, Chiba 277-8583, Japan}
\affiliation{Trans-Scale Quantum Science Institute, The University of Tokyo, Tokyo 113-0033, Japan}

\begin{abstract}

Parton showers which can efficiently incorporate quantum interference effects have been shown to be run efficiently on quantum computers. 
However, so far these quantum parton showers did not include the full kinematical information required to reconstruct an event, which in classical parton showers requires the use of a veto algorithm. 
In this work, we show that adding one extra assumption about the discretization of the evolution variable allows to construct a quantum veto algorithm, which reproduces the full quantum interference in the event, and allows to include kinematical effects. 
We finally show that for certain initial states the quantum interference effects generated in this veto algorithm are classically tractable, such that an efficient classical algorithm can be devised. 

\end{abstract}

\maketitle

\section{Introduction}

With recent impressive advances in quantum technologies, 
it is a fascinating problem to study possible applications of quantum computers to high energy physics (HEP) (see~Ref. \cite{Bauer:2022hpo} for the review and more references).
There are many computational problems for which quantum computers exponentially outperform the best known classical algorithms, while the reverse is known to be false~\cite{nielsen2010quantum}\footnote{The correct statement is that complexity class BPP, namely the range of computational problems that a classical computer can determine probabilistically in polynomial time, is fully contained in the complexity class BQP, which is the range computational problems that a quantum computer can determine probabilistically in polynomial time.}. However, classical computers are likely to outperform quantum computers for computational problems without a clear quantum advantage. 
It is therefore a good strategy to separate problems in HEP into smaller parts, and to use quantum computers for those problems where a clear advantage can  be established
while leaving other more tractable problems to classical computers.

Several proposals have been made in HEP to identify such classically intractable problems. 
The first is to use effective field theory techniques to identify universal field theory quantities that are non-perturbative in nature, and then use the capabilities of quantum computers to simulate such field theories for their computation~\cite{Bauer:2021gup}.
Another proposal was found in parton shower algorithms, which simplify the calculation of extra particles radiated from highly virtual particles. 
A parton shower is based on factorization theorems separating soft, collinear and hard radiation from one another~\cite{Collins:1987pm,Sterman:1995fz,Jaffe:1996zw,Bauer:2002nz}.
In theories containing multiple flavors of fermions charged under the gauge group, quantum interference effects can arise in the radiation process. 
It was shown in~\cite{Bauer:2019qxa} that a new quantum parton shower (QPS) algorithm could efficiently take into account these quantum interference effects (see also Refs.~\cite{Bepari:2020xqi,Li:2021kcs,Bepari:2021kwv,Macaluso:2021ngq}). 
While there were no applications of these results mentioned in Ref.~\cite{Bauer:2019qxa},
it was later pointed out in Ref.~\cite{Chigusa:2022act} that quantum interference effect
can have dramatic consequences in some dark sector scenarios beyond the standard model,
and hence can in principle be observed in experiments. 

In~\cite{Bauer:2019qxa} the QPS was formulated entirely as a quantum algorithm, with all relevant information stored in quantum registers.
These registers do not have to be measured until the end of the shower, and the result of the measurement produced a single event with the correct probability distribution.
It was mentioned in Ref.~\cite{Bauer:2019qxa} and further discussed in Ref.~\cite{Deliyannis:2022uyh} that performing mid-circuit measurements and controlling future quantum operations on the results of these measurements can greatly enhance the efficiency of the quantum algorithm.
While controlling quantum operations on classical information dynamically is not possible currently on publicly accessible quantum computers, this is expected to change in the coming years.

In this work we go one step further and introduce a quantum veto algorithm which is significantly more efficient than the previous QPS algorithms proposed. 
We will call the resulting algorithm a quantum veto parton shower (QVPS).
As we will discuss in detail later, for this QVPS to reproduce the correct quantum entanglement will require one extra condition to be satisfied when digitizing the evolution variable of the parton shower, but this extra requirement can be fulfilled for most interesting cases.
The QVPS proposed is a true quantum-classical hybrid algorithm, going beyond the mid-circuit measurements included in the previous algorithm.
As will be discussed, classical variables are distributed according to an overestimated probability distribution, and the quantum state which holds the required interference information is created in a quantum circuit that depends on the classical variables chosen.

A major advantage of the QVPS is that it allows to include kinematical information into the QPS which was not considered in previous work.
This will open the door to eventually having fully realistic quantum shower algorithms, which can capture all necessary information to produce fully differential event samples.
We furthermore believe that our quantum-classical hybrid implementation of the veto procedure 
will be of wider interest when we apply Monte Carlo techniques in combination with quantum computers.

A very interesting result of the QVPS construction is that the resulting quantum circuit is simple enough that under certain assumptions of the initial state before the shower it can be efficiently simulated classically.
This implies that in these cases the quantum interference effects can be tracked in a classical simulation, which makes the resulting physics available on a much faster timescale.

The rest of this paper is organized as follows. 
In~\cref{sec:Toy_model} and \cref{sec:multi_step}, we explain the basic idea of the quantum veto procedure in a very simple toy model,
first for a single step in~\cref{sec:Toy_model} and then multiple steps in~\cref{sec:multi_step}.
This discussion will help explain the basic setup without any complications from the particle physics considerations of an explicit parton shower.
In~\cref{sec:kin_QPS} we then introduce more complete discussion of the 
QVPS algorithm, and show numerical results for a case where a classical simulation of the QVPS is possible. 
Our conclusions are presented in~\cref{sec:conclusions}.

%
\section{A quantum veto algorithm for a simple toy model} \label{sec:Toy_model}
%

In this section, we introduce the veto algorithm for a very simple toy model and explain how it can be implemented on a quantum circuit for application to the quantum Monte Carlo simulations.

\subsection{Definition of the toy model} \label{sec:Toy_model_def}

Consider a quantum state consisting of a single qubit, parameterized by an angle $\phi$
\begin{align}
  \stateReg & = \alpha_0 \ket{0} + \alpha_1 \ket{1} \nonumber\\
  & \equiv \cos\frac{\phi}{2} \ket{0} + \sin\frac{\phi}{2} \ket{1}
  \,,
  \label{eq:state_s}
\end{align}
where we have defined $\alpha_0 = \cos(\phi/2)$ and used $\alpha_0^2 + \alpha_1^2 = 1$. 
This state can also be characterized by its density matrix
\begin{align}
    \rho_\stateLabel = \sum_{q,q'=0,1} \alpha_q \alpha_{q'}  \ket{q} \bra{q'}
    \,.
\end{align}

Now consider a probability distribution $f(q, \bmc; z)$ in a variable $z$ that depends on the quantum state $\ket q$ as well as a set of parameters $\bmc$. 
Assume that the allowed range in $z$ also depends on $\ket{q}$ and $\bmc$, $V(q, \bmc) \equiv [z_{\rm min}(q, \bmc), z_{\rm max}(q, \bmc)]$ and define the function $f(q, \bmc; z)$ to vanish outside of this range.
One also defines the total emission probability as the integral over all values $z$ 
\begin{align}
    f_{\rm tot}\intpars \equiv \int \! {\rm d} z \, f\pars 
    \,,
\end{align}
as well as the no-emission probability
\begin{align}
    f_{\rm no}\intpars \equiv 1 - f_{\rm tot}\intpars 
    \,.
\end{align}
For reasons that will become clear later, we will also assume that the total emission probability is much smaller than $1$:
\begin{align}
\label{eq:ftot_limit}
    f_{\rm tot}\intpars \ll 1
    \,.
\end{align}

This probability distribution can now be used to add an emission to the quantum state $\stateReg$. 
The probability to obtain a given value $z$ for the quantum state $\stateReg$ will be given by
\begin{align}
\label{eq:p_emission}
    p\cpars = \sum_q \alpha_q^2 f\pars
    \,,
\end{align}
and the probability for no emission will be
\begin{align}
\label{eq:p_no_emission}
    p_{\rm no}\intcpars = \sum_q \alpha_q^2 f_{\rm no}\intpars
    \,.
\end{align}
The density matrix of the state after adding an emission (or not) is given by
\begin{align}
    \rho(\bmc) = \rho_{\rm no}(\bmc) + \int \! {\rm d} z \, \rho(\bmc; z)
    \,,
\end{align}
with
\begin{align}
\label{eq:rho_no}
  \rho_{\rm no}\intcpars &= \sum_{q,q'=0,1} \alpha_q \alpha_{q'} \, \sqrt{f_{\rm no}\intpars} \sqrt{f_{\rm no}(q',\bmc)} \ket{q} \bra{q'}
  \,,
\end{align}
and
\begin{align}
\label{eq:rho_emission}
  \rho\cpars &= \sum_{q,q'=0,1} \alpha_q \alpha_{q'} \, \sqrt{f\pars} \sqrt{f(q',\bmc; z)} \ket{q} \bra{q'}
  \,.
\end{align}

Consider how one might computationally create this density matrix. 
If one could sample from the function $p(\bmc; z)$, one could choose if an emission happens, and if yes with what value of $z$ probabilistically, and then compute the corresponding density matrix. 
This is only possible if one can obtain an analytical expression of the inverse of the integrated function
\begin{align}
    {\cal P}(\bmc; z) \equiv \int_0^z {\rm d} z' \, p(\bmc; z')
    \,.
\end{align}
Given that $p(\bmc; z)$ is a weighted sum of different probability distributions, satisfying this condition is in general not possible, even if one can analytically sample from each function $f(q, \bmc; z)$.

A second option is to decompose the allowed range of $z$ into $\Ndisc$ independent pieces and sample one of them using the corresponding discrete probability distribution.
The computational complexity can easily be seen to scale linearly with $\Ndisc$, and in particular goes to infinity as one takes the continuum limit $\Ndisc \to \infty$. 

In the next section we show how one can use a veto algorithm to generate the density matrix for continuous values of $z$.

\subsection{Using a veto algorithm} \label{sec:Toy_model_veto}

The above toy model can be solved using a relatively straightforward modification of the well-known veto algorithm. 
In particular, one chooses an overestimate to the probability distribution that works for all values of the quantum parameter $q$
\begin{align}
  \over{f}\cpars \geq f\pars \qquad (\forall q)
  \,.
\end{align}
Also defining an overestimated phase space $\over{V} \supseteq V\labels$, one obtains the overestimated emission probability as
\begin{align}
  \over{f}_{\rm tot} \intcpars \equiv
  \int_{\over{V}} dz\, \over{f}\cpars
  \,,
\end{align}
which only depends on the classical parameters $\bmc$.
For later convenience, we assume the following relation
\begin{align}
\label{eq:ftot_over_limit}
  \over{f}_{\mathrm{tot}} (\bm{c}) \ll 1 \,,
\end{align}
which is in principle possible due to \cref{eq:ftot_limit}.

Choosing a well-behaved function, one can sample from this overestimated probability distribution by selecting a random number $\mathcal{R}$ from a uniformly distributed range $[0, 1)$. 
A value $\mathcal{R} > \over{f}_{\mathrm{tot}} (\bm{c})$ results in no emission, while for the remaining values one solves the equation
\begin{align}
    z = (\over{\mathcal{F}}_{\bm{c}})^{-1}(\mathcal{R})
\end{align}
with
\begin{align}
  \over{{\cal F}}_{\bm{c}}(z) \equiv \int_{\over{V}_{\leq z}} {\rm d}z'\, \over{\hat f}(\bmc;z') \,,
\end{align}
where $\over{V}_{\leq z} \equiv \over{V} \cap (-\infty, z]$ and 
the probability distribution 
$\over{\hat f}\cpars$ is defined by
\begin{align}
\label{eq:jsample}
    \over{\hat f}\cpars \equiv \frac{\over f\cpars}{\over {f}_{\rm tot}(\bmc)}
    \,.
\end{align}
Note that this requires that we can invert the function $\over{{\cal F}}_{\bm{c}}(z)$.
This selects no emission or a value of $v$ based on the overestimated probability distribution.

Next, this result is corrected by vetoing, i.e., discarding the sampled values with a certain probability. 
This veto procedure depends on the value of the quantum parameter $q$. 
More concretely, if an emission with value $z$ was selected in the previous step one accepts this event if and only if
\begin{align}
    z \in V\intpars\,, 
    \qquad {\rm and} \qquad \displaystyle{\frac{f\pars}{\over{f}\cpars} > \mathcal{R}'}
    \,,
\end{align}    
where $\mathcal{R}'$ is a new uniformly distributed random variable. 
If these conditions are not satisfied, we discard $z$ and add the event to the ones with no emission.
One can easily see that, after the vetoing procedure, the sampled values obey the probability distribution $f\pars$ as required.

Given that the parameter $q$ is a quantum variable, the veto procedure has to be performed using conditional gates in a quantum circuit. 
This will be discussed in the next section.

%
\subsection{Quantum circuit for veto procedure}
\label{subsec:QC_veto}
%

Consider the following quantum circuit.
\ifthenelse{\boolean{compileAll}}{
\begin{figure}[htp]
  \centering
  \begin{adjustbox}{width=\hsize}
    \begin{tikzcd}[column sep=0.2cm]
      \lstick{$\stateReg$} & \qw & \qw & \qw & \octrl{1} & \qw & \ctrl{1} & \qw & \qw & \qw \\
      \lstick{$\ket{e}$} & \qw & \qw & \qw & \gate{R_Y(\theta_0)} & \qw & \gate{R_Y(\theta_1)} & \qw & \meter{} & \qw\\
      \lstick{$\bm{c}$} & \cw & \cwbend{1} & \cw & \cwbend{-1} & \cw & \cwbend{-1} & \cw & \cw & \cw\\
      \lstick{$z$} & \cw & \gate[cwires={1}]{\over{\hat f}\cpars} & \cw & \cwbend{-1} & \cw & \cwbend{-1} & \cw & \cw & \cw
     \end{tikzcd}
  \end{adjustbox}
  \caption{
    An example quantum circuit for the veto procedure. 
  }
  \label{fig:veto_example}
\end{figure}
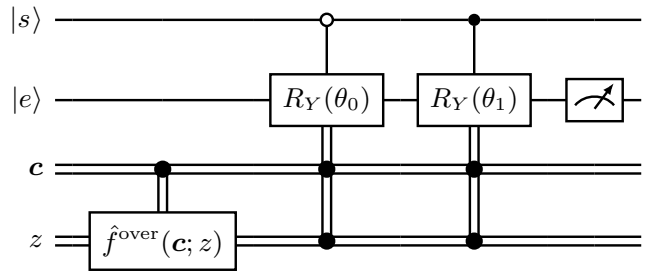
}{\textbf{[Figure omitted to avoid too long compile time]}}
We assume that the qubit $\stateReg$ is already in the state given by \cref{eq:state_s}, while the qubit $\ket{e}$ is assumed to be in the initial state $\ket{0}$. 
The classical wire labeled by $\bm{c}$ is assumed to hold the values of the parameters $\bm{c}$. 
The other classical wire labeled by $z$ holds the value of $z$, which is sampled according to the probability distribution in \cref{eq:jsample}.
This is indicated by the classical ``gate'' $\over {\hat f}\cpars$.

The $R_Y$ rotation gates on the register $\ket{e}$ are controlled on the qubit $\stateReg$. 
The rotation angles are given by
\begin{align}
    \theta_{q} &= \theta_{q}(q, \bmc; z) \nonumber\\
    & = \hat\Theta\left({f}(q, \bmc;z)/\over{\hat{f}}(\bmc;z)\right)
    \,,
\end{align}
where
\begin{align}
    \hat\Theta(x) \equiv 2 \arctan \sqrt{ \frac{x}{1-x} }
    \,.
\end{align}

For each value of $z$ one can work out the result of this quantum circuit.
One can show very easily that for a given value $z$ of the classical register the state before the measurement of the $\ket{e}$ register is
\begin{widetext}
\begin{align}
  \ket{\psi(z)} = \sum_q \alpha_q \ket{q} \Biggl(&
    \sqrt{1- \frac{f(q, \bmc;z)}{\over{\hat f}(\bmc;z)}} \ket{0_e} 
    + \sqrt{\frac{f(q,\bmc;z)}{\over{\hat f}(\bmc;z)}} \ket{1_e}
  \Biggr) 
  \,.
  \label{eq:state_veto}
\end{align}
After measuring the register $\ket{e}$ density matrix of the system becomes
\begin{align}
    \rho(z) =& \sum_{q,q'} \alpha_q \alpha_{q'} \ket{q}\bra{q'}
    \Biggl(
    \sqrt{1- \frac{f(q, \bmc;z)}{\over{\hat f}(\bmc;z)}} \sqrt{1- \frac{f(q', \bmc;z)}{\over{\hat f}(\bmc;z)}} \ket{0_e} \bra{0_e}
    + 
    \sqrt{\frac{f(q',\bmc;z)}{\over{\hat f}(\bmc;z)}}\sqrt{\frac{f(q,\bmc;z)}{\over{\hat f}(\bmc;z)}} \ket{1_e}\bra{1_e}
  \Biggr)\,.
\end{align}
Importantly, the second term reproduces the correct density matrix after the emission \cref{eq:rho_emission} with an appropriate normalization factor.

Each of the values $z$ is obtained with probability $\over{\hat f}(\bmc; z)$ defined in \cref{eq:jsample}.
This implies that the ensemble average of the density matrix, describing the average state evolution when we repeat the simulation multiple times, is given by
\begin{align}
    \rho^{\rm veto}(\bmc) &= \int \! {\rm d} z \,  \over{\hat f}(\bmc; z) \rho(z)\nonumber\\
    &= \rho^{\rm veto}_{\rm no}(\bmc) \ket{0_e} \bra{0_e} + \int \! {\rm d}z\,  \rho^{\rm veto}(\bmc; z)\ket{1_e} \bra{1_e}\,.
\end{align}
Using \cref{eq:ftot_over_limit} one obtains 
\begin{align}
    \rho_{\rm no}^{\rm veto}(\bmc) 
     &\simeq \int \! {\rm d} z \,  \over{\hat f}(\bmc; z) \left[\sum_{q=0,1}\alpha_q^2  \left(1- \frac{f(q,\bmc;z)}{\over{\hat f}(\bmc; z)}\right)\ket{q}\bra{q}
     +
     \alpha_0 \alpha_1 \left( 1 - \frac{1}{2} \sum_{q=0,1}\frac{f(q,\bmc;z)}{\over{\hat f}(\bmc; z)}\right)
    \left( \ket{0}\bra{1} + \ket{1}\bra{0}\right)\right]
    \nonumber\\
     &= \sum_{q=0,1}\alpha_q^2  \left(1- f_{\rm tot}(q, \bmc)\right)\ket{q}\bra{q}
     + \alpha_0 \alpha_1 \left( 1 - \frac{1}{2} \sum_{q=0,1}f_{\rm tot}(q,\bmc)\right)
    \left( \ket{0}\bra{1} + \ket{1}\bra{0}\right)
    \nonumber\\
     &\simeq \sum_{q=0,1}\alpha_q^2  \left(1- f_{\rm tot}(q, \bmc)\right)\ket{q}\bra{q}
     + \alpha_0 \alpha_1 \left( 1 - \frac{1}{2} f_{\rm tot}(0,\bmc)\right)\left( 1 - \frac{1}{2} f_{\rm tot}(1,\bmc)\right)
    \left( \ket{0}\bra{1} + \ket{1}\bra{0}\right)
    \nonumber\\
     &\simeq \sum_{q=0,1}\alpha_q^2  \left(1- f_{\rm tot}(q, \bmc)\right)\ket{q}\bra{q}
     + \alpha_0 \alpha_1 \sqrt{1- f_{\rm tot}(0, \bmc)}\sqrt{1 - f_{\rm tot}(1, \bmc)} 
    \left( \ket{0}\bra{1} + \ket{1}\bra{0}\right)
    \nonumber\\
    &= \sum_{q, q'}\alpha_q^2  \sqrt{f_{\rm no}(q, \bmc)}\sqrt{f_{\rm no}(q', \bmc)}\ket{q}\bra{q'}
    \,.
\end{align}
\end{widetext}
Thus, the expression reproduces the non-emission density matrix \cref{eq:rho_no} up to terms of $\mathcal{O}\left((\over{f}_{\mathrm{tot}}(q, \bmc))^2\right)$.
It is the presence of these correction terms that requires the condition~\cref{eq:ftot_limit}.

We can test this procedure by performing a simulation for a simple probability distribution.
We choose
\begin{align}
    f(q,\bm{c}; z) = 2 \, c_1 \, z \, \delta_{q, 0}
    + 3 \, c_2 \, z^2 \, \delta_{q, 1}
    \,,
\end{align}
with $0 \leq c_1, c_2 \leq 1$ and a $q$-independent phase space $z \in [0,1]$. 
We use the overestimated quantities
\begin{align}
    \over{f}(z) = \max (2c_1, 3c_2) \,,
\end{align}
and $\over{V} = [0,1]$, which is valid only for $c_1, c_2 \ll 1$.
This gives 
\begin{align}
    \over{f}_{\mathrm{tot}} &= \max (2c_1, 3c_2)
    \,, \\
    \over{\hat{f}} (z) &= 1
    \,.
\end{align}
For our numerical results we choose $c_1=c_2=0.25$, and an initial state with $\alpha_0^{\mathrm{init}} = \alpha_1^{\mathrm{init}} = 1/\sqrt{2}$. 

\begin{figure}[th]
  \centering
  \includegraphics[width=\hsize]{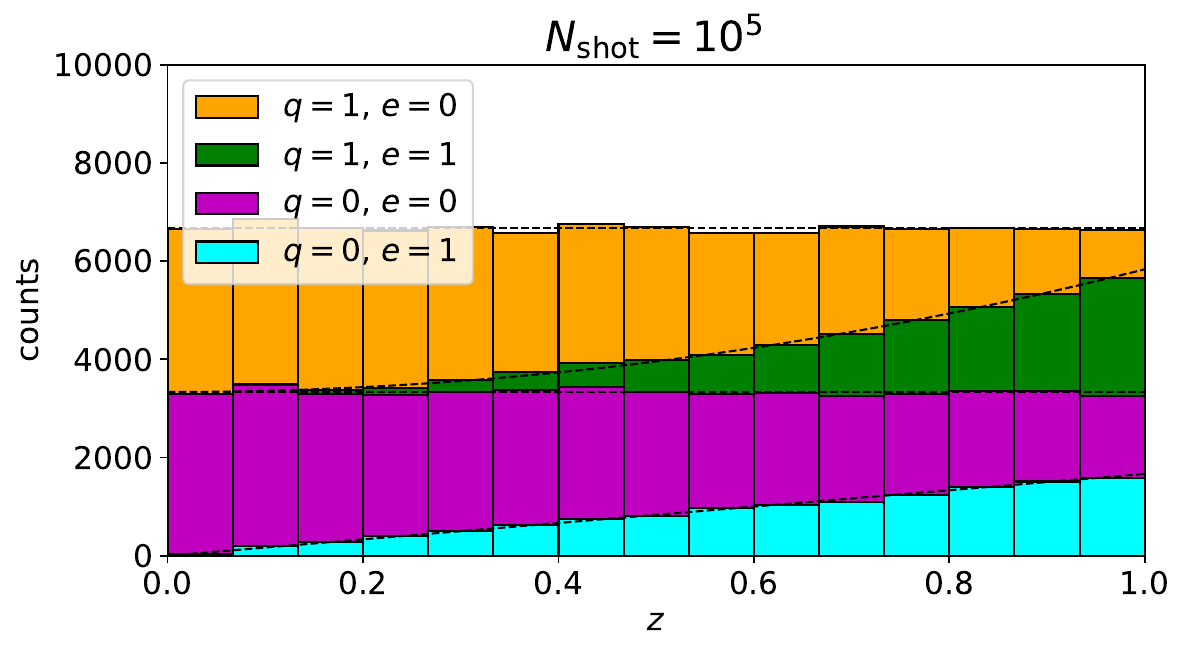}
  \caption{
    Simulation result of the quantum circuit given in \cref{fig:veto_example}.
    The cyan histogram corresponds to the distribution of $z$ when $e=1$, which correctly reproduces the probability distribution $f(q ; z)$ under the setup described in the text.
  }  
  \label{fig:veto_res1}
\end{figure}
In \cref{fig:veto_res1}, we show a simulation result with the number of shots $N_{\mathrm{shot}} = 10^5$.
We measure both the state $\stateReg$ and the emission register $\ket{e}$ after the simulation, and the result is shown in terms of a stacked histogram.
The cyan (green) histogram shows the result with $q = 0$ ($q = 1$) and $e = 1$, therefore giving the distribution in $z$, while the magenta (orange) one shows the same for $q = 0$ ($q = 1$) but with the non-emission result $e = 0$. 
Three dashed lines represent the analytical expectation of the probability distributions.
We can see that the simulation result agrees with those expectations within the statistical errors.
In particular, the cyan and green results correctly reproduce the probability distribution $f(q=0, \bm{c} ; z)$ and $f(q=1, \bm{c} ; z)$, respectively, weighted according to the input state $\ket{s}$.

\begin{figure}[th]
  \centering
  \includegraphics[width=\hsize]{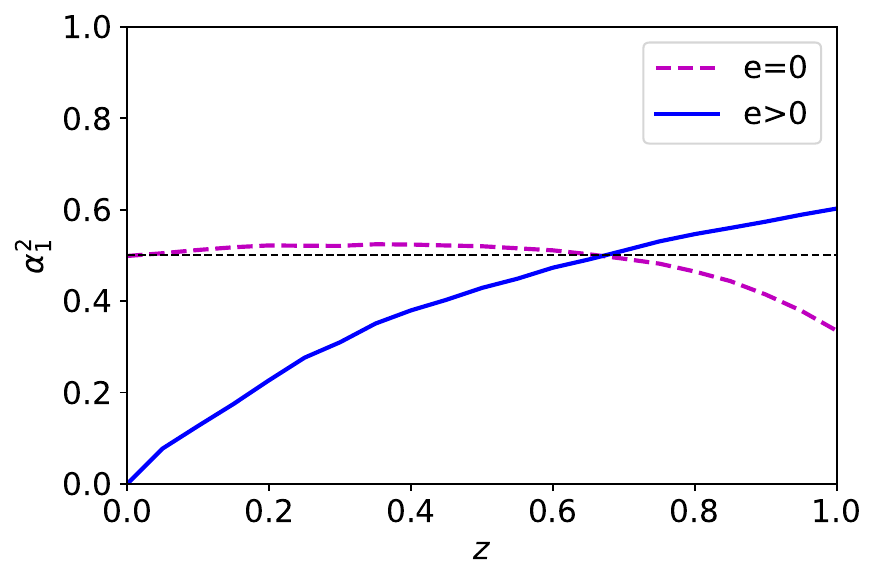}
  \caption{Dependence of the quantum state after the simulation on the classically sampled value of $z$.}
  \label{fig:veto_res2}
\end{figure}

We also show how the quantum state stored in the qubit $\stateReg$ changes after the simulation in Fig.~\ref{fig:veto_res2}.
The $x$-axis of the plot denotes the classically sampled value of $z$, while the $y$-axis the probability $\alpha_1^2$ of finding $q=1$ after the simulation.
The purple dashed (blue solid) line shows the result without (with) emission.
This plot clearly shows that $\alpha_1^2$ after the simulation is different from its initial value, $\left( \alpha_1^{\mathrm{init}} \right)^2 = 0.5$, and depends on $z$ and $e$.
It demonstrates that the simulation result can affect the quantum state, stressing the importance of performing the simulation quantum mechanically.

\subsection{Adding additional qubits to \texorpdfstring{$\stateReg$}{|s>}}
\label{sec:gate_count}
The quantum algorithm discussed so far contained only a single qubit with $q=0,1$.
An extension to multiple qubits is straightforward, but the total number of probability distributions is given by the total number of quantum states, so for $n_q$ qubits that number is $2^{n_q}$. 
In the algorithm discussed so far, each of these probability distributions required its own rotation on the register $\ket{e}$, such that the number of gates required scales exponentially with the number of qubits.

As we will now show, the number of gates can be reduced to scale linear with the number of qubits for an important special case of the probability density function. 
In this special case, one has $n_q$ different distributions
\begin{align}
  g_k(q_k, \bmc_k ; z)
  \,,
  \label{eq:f_LC}
\end{align}
with $k=1,\dots,n_q$.
Here $q_k=0,1$ represents the value of one of the $n_q$ qubits denoted as $\stateReg[k]$, $\bmc_k$ a set of parameters, and for each $k$ one can in principle have a different functional form $g_k$.
We furthermore assume that states generated from different distributions do not interfere with one another. 
This can be written by introducing the operator
\begin{align}
\label{eq:splitting_op}
    f(\bmq, \bmc; z) &=  f_{\rm no}(\bmq, \bmc) \ket{0_k}\bra{0_k}\nonumber\\
    & + \sum_{k=1}^{n_q} g_f(q_k, \bmc_k ; z) \ket{k} \bra{k}
    \,,
\end{align}
where $\bmq$ denotes the set of all qubit values, while $\bmc$ denotes the set of all classical parameters.
$\ket{0_k}$ and $\ket{k}$ are states of a fictitious register that preserves from which $g_k$ the emission occurs (or not).
The projector onto these states ensures that there is no interference between the distributions with different values of $k$.
We have also introduced 
\begin{align}
    f_{\rm no}(\bmq, \bmc) \equiv 1 - f_{{\rm tot}}(\bmq, \bmc)
    \,,
\end{align}
with
\begin{align}
    f_{{\rm tot}}(\bmq, \bmc) \equiv \sum_k \int\! {\rm d}z \, g_k(q_k, \bmc_k ; z)
    \,.
\end{align}

One finds the probability of having no emission to be
\begin{align}
    p_{\rm no}(\bmc) = \sum_{\bmq} \alpha_q f_{\rm no}(\bmq, \bmc)
    \,,
\end{align}
while the probability of having an emission with given $k$ and $z$ is given by
\begin{align}
    p_k(\bmc) = \sum_{\bmq} \alpha_q g_k(\bmq, \bmc; z)
    \,.
\end{align}

The density matrix can now be written as
\begin{align}
\label{eq:density_1}
        \rho(\bmc) = \rho_{\rm no}(\bmc) + \sum_k \int \! {\rm d} z \, \rho_k(\bmc; z)
    \,,
\end{align}
with
\begin{align}
\label{eq:density_2}
      \rho_{\rm no}\intcpars = & \sum_{\bmq,\bmq'} \alpha_{\bmq} \alpha_{\bmq'} \, \sqrt{f_{\rm no}(\bmq, \bmc)} \sqrt{f_{\rm no}(\bmq',\bmc)} 
      \nonumber\\
      & \times \ket{\bmq} \bra{\bmq'}
  \,,\notag\\
      \rho_k(\bmc; z) = & \sum_{\bmq,\bmq'} \alpha_{\bmq} \alpha_{\bmq'} \sqrt{g_k(q_k, \bmc_k; z)} \sqrt{g_k(q'_k, \bmc_k; z)}
      \nonumber\\
      & \times \ket{\bmq} \bra{\bmq'}
  \,,
\end{align}
where $\ket{\bmq} \equiv \ket{q_1} \otimes\cdots \otimes\ket{q_n}$ and $\alpha_{\bmq} \equiv \Braket{\bmq | {\bm \stateLabel}}$ with $\ket{\bm \stateLabel}$ being the $n_{\bmq}$-qubit state register.
We can now construct an overestimated probability distribution for each $k$, such that
\begin{align}
    \over g_k(\bmc_k; z) \geq g_k(q_k, \bmc_k ; z)
    \,.
\end{align}
We also define 
\begin{align}
    \over g_{k,{\rm tot}}(\bmc_k) \equiv \int \! {\rm d}z \, \over g_k(\bmc_k; z) 
    \,,
\end{align}
as well as the normalized expressions
\begin{align}
\label{eq:jfprobs}
    \over {\hat g}_k(\bmc_k; z) &\equiv \frac{\over {g}_k(\bmc_k; z)}{\over g_{k,{\rm tot}}(\bmc_k)}
    \,,\nonumber\\
    \over {\hat g}_{k,\mathrm{tot}}(\bmc) &\equiv \frac{\over {g}_{k,{\rm tot}}(\bmc_k)}{\sum_k \over {g}_{k,\mathrm{tot}}(\bmc_k) }
    \,.
\end{align}

The veto algorithm is very similar to the one obtained in Section~\ref{subsec:QC_veto}, with a slight modification to the original classical sampling. 
In particular, in addition to the classical variable $z$, one needs an additional classical variable $k$, which labels the term in the sum \cref{eq:f_LC}, and therefore also the qubit $\stateReg[k]$. 
The veto algorithm now proceeds as follows:
\begin{enumerate}
    \item Choose an integer $k$ according to the normalized probabilities $\over {\hat g}_{k,\mathrm{tot}}(\bmc)$
    \item Run the algorithm of Section~\ref{subsec:QC_veto} using $\stateReg = \stateReg[k]$ and rotation angle in the $R_Y$ gate which depends on the classical values $z$ (determined in the first step of this algorithm) and the value $k$ sampled in the first step
    \begin{align}
        \theta_{q} &= \theta_{q}(k, \bmq, \bmc; z) \nonumber\\
    & = \hat \Theta \left( \frac{{g_k}(q_k, \bmc_k;z)}{\over {\hat g}_{k,{\rm tot}}(\bmc)\over{\hat{g}}_{k}(\bmc_k; z)}\right)
    \,.
    \end{align}
\end{enumerate}
\ifthenelse{\boolean{compileAll}}{
\begin{figure}[htp]
  \centering
  \begin{adjustbox}{width=\hsize}
    \begin{tikzcd}[column sep=0.2cm]
      \lstick{$\stateReg[1]$} & \qw & \qw & \qw & \qw & \qw & \qw & \qw & \qw & \qw & \qw \\
      \lstick{$\vdots~~$} &&&&&&&&&& \\
      \lstick{$\stateReg[k]$} & \qw & \qw & \qw & \qw & \octrl{3} & \qw & \ctrl{3} & \qw & \qw & \qw \\
      \lstick{$\vdots~~$} &&&&&&&&&& \\
      \lstick{$\stateReg[n_q]$} & \qw & \qw & \qw & \qw & \qw & \qw & \qw & \qw & \qw & \qw \\
      \lstick{$\ket{e}$} & \qw & \qw & \qw & \qw & \gate{R_Y^{(j)}(\theta_0)} & \qw & \gate{R_Y^{(j)}(\theta_1)} & \qw & \meter{} & \qw\\
      \lstick{$\bmc$} & \cw & \cwbend{2} & \cwbend{1} & \cw & \cwbend{-1} & \cw & \cwbend{-1} & \cw & \cw & \cw\\
      \lstick{$z$} & \cw & \cw & \gate[cwires={1}]{\over {\hat g}_k(\bmc_k; z)} & \cw & \cwbend{-1} & \cw & \cwbend{-1} & \cw & \cw & \cw\\
      \lstick{$k$} & \cw & \gate[cwires={1}]{\over {\hat g}_{k,{\rm tot}}(\bmc)} & \cw & \cw & \cwbend{-1} & \cw & \cwbend{-1} & \cw & \cw & \cw
     \end{tikzcd}
  \end{adjustbox}
  \caption{
    A single step of a quantum veto algorithm according to~\cref{eq:splitting_op}. Note that each step will in general have a different classical value $k$, and will therefore be controlled on a different qubit $\ket{q_k}$. 
  }
  \label{fig:veto_example_fullq}
\end{figure}
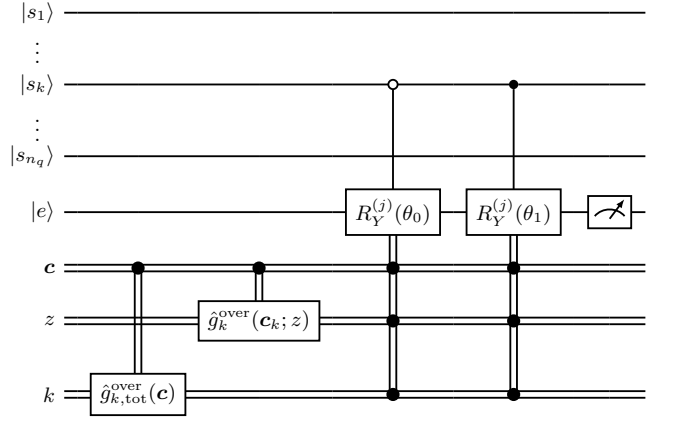
}{\textbf{[Figure omitted to avoid too long compile time]}}

For each value of $k$ and $z$ one can work out the result of this quantum circuit.
This step is almost the same as in the previous case where only a single qubit was present in the state $\ket{q}$, and the only modifications are keeping track of the integer $k$.
The state before the measurement of the $\ket{e}$ register depends on the values of $k$ and $z$, one finds very similar to the result in~\cref{eq:state_veto}
\begin{align}
  \psi_k(z) = \alpha_{q} \ket{q} \Biggl(&
    \sqrt{1- \frac{{g_k}(q_k, \bmc_k;z)}{\over {\hat g}_{k,{\rm tot}}(\bmc)\over{\hat{g}}_{k}(\bmc_k; z)}} \ket{0_e} 
    \notag \\
    &
    + \sqrt{\frac{{g_k}(q_k, \bmc_k;z)}{\over {\hat g}_{k,{\rm tot}}(\bmc)\over{\hat{g}}_{k}(\bmc_k; z)}} \ket{1_e}
  \Biggr) 
  \,.
  \label{eq:state_veto_multiple}
\end{align}
Measuring the register $\ket{e}$ and taking into account the probabilities~\cref{eq:jfprobs} with which each value of $k$ and $z$ are generated classically, the total density matrix of the circuit is given by
\begin{align}
    \rho^{\rm veto}(\bmc) 
    &= \rho^{\rm veto}_{\rm no}(\bmc) \ket{0_e} \bra{0_e} 
    \nonumber\\
    & \qquad + \sum_{k} \int \! {\rm d} z \, \rho_k^{\rm veto}(\bmc_k; z)\ket{1_e} \bra{1_e}
    \,.
\end{align}
Following similar steps as in the previous section, one obtains 
\begin{align}
    \rho_{\rm no}^{\rm veto}(\bmc) 
     &\simeq \sum_{\bmq,\bmq'} \alpha_{\bmq} \alpha_{\bmq'} \, \sqrt{f_{\rm no}(\bmq, \bmc)}\sqrt{f_{\rm no}(\bmq', \bmc)}\ket{\bmq}\bra{\bmq'}
     \,,\notag\\
    \rho_{k}^{\rm veto}(\bmc_k; z) 
     &= \sum_{\bmq,\bmq'} \alpha_{\bmq} \alpha_{\bmq'} \, \sqrt{{g_k}(q_k, \bmc_k;z)}
     \notag\\
     & \qquad \times \sqrt{{g_k}(q'_k, \bmc_k;z)}\ket{\bmq}\bra{\bmq'}
     \,,
\end{align}
reproducing the results of~\cref{eq:density_2}. 

Note that in order to derive these expressions, we needed to use the generalization of~\cref{eq:ftot_over_limit}, namely
\begin{align}
  \label{eq:cond_over_fullq}
    \sum_{k=1}^{n_q} {\over{g}_{k , {\rm tot}}}(\bmc_k) \ll 1
    \,,
\end{align}
One therefore needs to be careful with these expressions in the limit $n_q \to \infty$, which will be discussed more when we discuss an actual parton shower.

\subsection{Simulating the quantum veto algorithm classically}
\label{sec:classical_simulation}
The quantum state $\ket{\bm \stateLabel}$ is determined by $2^{n_q}-1$ independent complex numbers $\alpha_{\bmq}$.
As can be seen from~\cref{fig:veto_example_fullq}, the veto algorithm of~\cref{sec:gate_count} only contains a quantum gate acting on the qubit $\ket{e}$ and which is controlled by a single qubit $\stateReg[k]$, where the value of $k$ was determined classically.
The resulting quantum state is given in~\cref{eq:state_veto_multiple} and is obtained by multiplying the complex numbers $\alpha_{\bmq}$ by a number that only depends on the value of $q_k$. 

If the original quantum state 
\begin{align}
\label{eq:state_orig}
    \psi = \sum_{\bmq} \alpha_{\bmq} \ket{\bmq}
\end{align}
is a direct product of the individual qubits $\stateReg[k]$ one can write
\begin{align}
    \stateReg[k] &= \sum_{q_k=0,1} \alpha_{k,q_k} \ket{q_k}
    \,,\notag\\
    \alpha_{\bmq} &= \alpha_{1,q_1} \times \cdots \times \alpha_{n,q_n}
    \,.
\end{align}
In this case, the action of the quantum circuit~\cref{fig:veto_example_fullq} can be worked out very easily, and one finds that the state $\ket{e}$ is measured to be in the $\ket{0_e}$ and $\ket{1_e}$ with probabilities
\begin{align}
    p_k(z;0_e) &= \sum_{q_k} \left|\alpha_{k,q_k}\right|^2 \left(1- \frac{{g_k}(q_k, \bmc_k;z)}{\over {\hat g}_{k,{\rm tot}}(\bmc)\over{\hat{g}}_k(\bmc_k; z)}\right)
    \,,\notag\\
    p_k(z;1_e) &= \sum_{q_k} \left|\alpha_{k,q_k}\right|^2  \frac{{g_k}(q_k, \bmc_k;z)}{\over {\hat g}_{k,{\rm tot}}(\bmc)\over{\hat{g}}_{k}(\bmc_k; z)}
    \,.
\end{align}
Note that this probability only contains two terms in the sum, and can therefore be computed easily classically.
Once the value of $\ket{e}$ has been measured, the resulting state is given by 
\begin{align}
    \psi_k(z;e) = \sum_{\bmq} \alpha_{1,q_1} \times \cdots \alpha_{k,q_k}(z; e) \times \cdots \alpha_{n,q_n} \ket{\bmq}
    \,,
\end{align}
with $e = 0,1$, where
\begin{align}
    \alpha_{k,q_k}(z;0_e) &= \frac{\alpha_{k,q_k}}{\sqrt{p_k(z;0_e)}} \sqrt{1- \frac{{g_k}(q_k, \bmc_k;z)}{\over {\hat g}_{k,{\rm tot}}(\bmc)\over{\hat{g}}_k(\bmc_k; z)}}
    \,,\notag\\
    \alpha_{k,q_k}(z;1_e) &= \frac{\alpha_{k,q_k}}{\sqrt{p_k(z;1_e)}} \sqrt{\frac{{g_k}(q_k, \bmc_k;z)}{\over {\hat g}_{k,{\rm tot}}(\bmc)\over{\hat{g}}_k(\bmc_k; z)}}
    \,.
\end{align}
Thus, the two possible states differ from the original state given in~\cref{eq:state_orig} only by the two complex numbers specifying the state of qubit $\stateReg[k]$, which can again be tracked easily classically.

For more complex initial states that contain $n_q'$ entangled qubits, the complexity of the classical calculations scales exponentially in $n_q'$. 
This implies that the classical simulation of the quantum veto algorithm is only possible for initial states with low entanglement.
We will discuss the implications of this for a QVPS in the coming section.

\section{Multi-step quantum veto simulation} \label{sec:multi_step}

So far, we have focused only on the single-step simulation to determine whether the emission occurs or not.
The procedure can easily be extended to the multi-step simulation with possible multi-emissions.
As we will see below, this extension introduces an interesting quantum interference effect.

Consider an $n_q$ qubit-state toy model with $\ket{\bm{\stateLabel}}$, together with a set of probability distributions $f^{(i)}(q, \bmc^{(i)}; z)$ together with $f^{(i)}_{{\rm tot}}(q, \bmc^{(i)}) = \int \! {\rm d} z f^{(i)}(q, \bmc^{(i)}; z)$.
Each of the different values of $i = 0 \ldots N$ denotes a step, and at each step $i$ an emission occurs with probability $f^{(i)}_{{\rm tot}}(q, \bmc^{(i)})$ and value $z^{(i)}$ determined by $f^{(i)}(q, \bmc^{(i)}; z^{(i)})$. 
As before, the probability distribution is a sum over different distributions, each acting on a different qubit in the register $\ket{\bm{\stateLabel}}$
\begin{align}
    f^{(i)}(q, \bmc^{(i)}; z^{(i)}) = \sum_k g_k^{(i)}(q_k, \bmc_k^{(i)}; z^{(i)})
    \,.
\end{align}

To keep track of the classical information generated at previous steps, we define a classical history register $\bmh^{(i)}$. 
It holds all classical information about emissions up to the step $i$, so $\bmh^{(i)} = \bigcup_{j< i} \left\{ e^{(j)}, z^{(j)}, k^{(j)} \right\}$. 
We also assume that the classical parameters depend on the history 
\begin{align}
    \bmc^{(i)} \equiv \bmc(\bmh^{(i)})
    \,,
\end{align}
and that at the beginning of each step the quantum state is changed as
\begin{align}
    \ket{\bm{\stateLabel}} \to U^{(i)}\ket{\bm{\stateLabel}}
    \,,
\end{align}
where the unitary transformation again depends on the classical history
\begin{align}
    U^{(i)} \equiv U(\bmh^{(i)})
    \,.
\end{align}
We assume that after the unitary transformation at the beginning of the first step the quantum state is given by \cref{eq:state_s}. 

We will now consider two emissions in sequence. 
The two possible classical histories at the beginning of the second emission are that no emission occurred (denoted by $\bmh^{(2)}_{\rm no}$), or that an emission occurred (denoted by $\bmh^{(2)}_{\rm em}$). 
After the $2$-step simulation ($N=2$), the density matrix can be written as
\begin{align}
  \rho = \rho^{(0)} + \sum_k \int \! {\rm d} z \, \rho_k^{(1)}(z) + \sum_{k_1 k_2}\int \! {\rm d} z_0 \, {\rm d} z_1 \, \rho_{k_1 k_2}^{(2)}(z_0, z_1)
  \,.
\end{align}
The first term denotes the density matrix if no emission has happened after either of the two steps
\begin{widetext}
\begin{align}
  \rho^{(0)} =& \sum_{rr'} \alpha_r \alpha_{r'}  \, 
  \sum_{qq'} [U(\bmh^{(2)}_{\rm no})]_{qr} [U_2^*(\bmh^{(2)}{\rm no})]_{q' r'}
  \notag \\
  &\times \sqrt{f_{\rm no}(\bmr, \bmc^{(1)})} \sqrt{f_{\rm no}(\bmr',\bmc^{(1)})} \sqrt{f_{\rm no}(\bmq, \bmc(\bmh^{(2)}_{\rm no}))} \sqrt{f_{\rm no}(\bmq',\bmc(\bmh^{(2)}_{\rm no})} 
  \nonumber\\
  & \times \ket{\bmq} \bra{\bmq'} \ket{0_{k_1}}\bra{0_{k_1}} \ket{0_{k_2}}\bra{0_{k_2}}
  \,.
\end{align}
If there has been a single emission, this emission could have happened in the first or the second step, so we need to sum over both contributions
\begin{align}
    \rho_k^{(1)}(z) =& \sum_{k} \sum_{r,r'} \alpha_r \alpha_{r'} \,
  \sum_{qq'} [U_2(\bmh^{(2)}_{\rm em})]_{qr} [U_2^*(\bmh^{(2)}_{\rm em})]_{q' r'}
  \notag \\
  &\times \sqrt{g_{k}(r_{k}, \bmc_{k}^{(1)} ; z)} \sqrt{g_{k}(r_{k}, \bmc_{k}^{(1)} ; z)} \sqrt{f_{\rm no}(\bmq, \bmc(\bmh^{(2)}_{\rm em})))} \sqrt{f_{\rm no}(\bmq',\bmc(\bmh^{(2)}_{\rm em})))} 
  \ket{\bmq} \bra{\bmq'} 
  \notag \\
  +& \sum_{k} \sum_{r,r'} \alpha_r \alpha_{r'} \,
  \sum_{qq'} [U_2(\bmh^{(2)}_{\rm no})]_{qr} [U_2^*(\bmh^{(2)}_{\rm no})]_{q' r'}
  \notag \\
  &\times \sqrt{f_{\rm no}(\bmr, \bmc^{(1)})} \sqrt{f_{\rm no}(\bmr',\bmc^{(1)})} \sqrt{g_{k}(q_{k}, \bmc_{k}(\bmh^{(2)}_{\rm no}); z)} \sqrt{g_{k}(q'_{k}, \bmc_{k}(\bmh^{(2)}_{\rm no}); z)}
  \ket{\bmq} \bra{\bmq'}
  \,,
\end{align}
where the first term denotes the emission happening in the first step, while the second term has the emission happening in the second step.
Finally, we have the contribution where an emission happened in both steps, which gives
\begin{align}
  \rho^{(2)}_{k_1k_2}(z_1, z_2) =& \sum_{k_0, k_1} \sum_{r,r'} \alpha_r \alpha_{r'} \,
  \sum_{qq'} [U_2(\bmh^{(2)}_{\rm em})]_{qr} [U_2^*(\bmh^{(2)}_{\rm em})]_{q' r'}
  \notag \\
  &\times \sqrt{g_{k_1}(r_{k_1}, \bmc_{k_1}^{(1)} ; z_1)} \sqrt{g_{k_1}(r_{k_1}, \bmc_{k_1}^{(1)} ; z_1)} \sqrt{g_{k_2}(q_{k_2}, \bmc_{k_2}(\bmh^{(2)}_{\rm em}) ; z_2)} \sqrt{g_{k_2}(q_{k_2}, \bmc_{k_2}(\bmh^{(2)}_{\rm em}) ; z_2)}
  \ket{\bmq} \bra{\bmq'}
  \,.
\end{align}
\end{widetext}
As we have done in \cref{sec:gate_count}, we have introduced fictitious qubit registers $\ket{k_j}$ that preserve from which $g_{k_j}$ the emission occurs (or not) in the $j$-th step.

These expressions clearly demonstrate an important quantum interference effect.
To see this, it is sufficient to consider a simple example with $U_{2} (\bmh) = {\rm id}$ and having the same classical register $\bmc$, independent of the history.
The probabilities of having $0$, $1$, and $2$ emissions are then given by
\begin{align}
  p^{(0)}(\bmc) &= \sum_{\bmq} \alpha_{\bmq}^2 f_{\rm no}(\bmq, \bmc)^2
  \,,\\
  p^{(1)}_k(\bmc; z) &= \sum_{\bmq} 2\alpha_{\bmq}^2 f_{\rm no}(\bmq, \bmc) g_k(\bmq_k, \bmc_k; z)
  \,,\notag\\
  p^{(2)}_{k_1k_2}(\bmc; z_1, z_2) &= \sum_{\bmq} \alpha_{\bmq}^2 g_{k_1}(\bmq_{k_1}, \bmc_{k_1}; z_1) g_{k_2}(\bmq_{k_2}, \bmc_{k_2}; z_2)
  \,,\notag
\end{align}
respectively, which should be compared with products of individual (non-)emission probability as
\begin{align}
  p^{(0)}(\bmc) &\neq p_{\mathrm{no}}(\bmc)^2
  \,,\\
  p^{(1)}_k(\bmc; z) &\neq 2 p_{\mathrm{no}}(\bmc) p_k(\bmc; z)
  \,,\\
  p^{(2)}_{k_1k_2}(\bmc; z_1, z_2) &\neq p_{k_1}(\bmc; z_1) p_{k_2}(\bmc; z_2)
  \,,
\end{align}
with $p_k(\bmc; z)$ and $p_{\mathrm{no}}(\bmc)$ are given in \cref{eq:density_2}.

\ifthenelse{\boolean{compileAll}}{
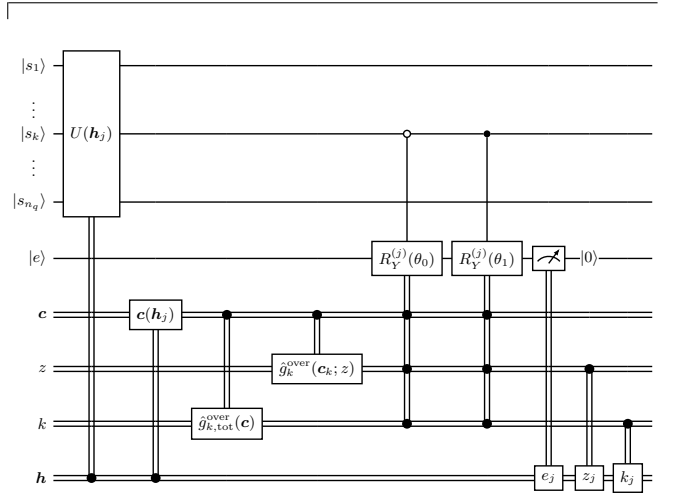
\begin{figure}[htp]
  \centering
  \begin{adjustbox}{width=\hsize}
    \begin{tikzcd}[column sep=0.2cm]
      \lstick{$\stateReg[1]$} & \gate[5,nwires={2,4}]{U(\bmh_j)} & \qw & \qw & \qw & \qw & \qw & \qw & \qw & \qw & \qw \\
      \lstick{$\vdots~~$} \\
      \lstick{$\stateReg[k]$} && \qw & \qw & \qw & \octrl{3} & \ctrl{3} & \qw & \qw & \qw & \qw \\
      \lstick{$\vdots~~$} \\
      \lstick{$\stateReg[n_q]$} && \qw & \qw & \qw & \qw & \qw & \qw & \qw & \qw & \qw \\
      \lstick{$\ket{e}$} & \qw & \qw & \qw & \qw & \gate{R_Y^{(j)}(\theta_0)} & \gate{R_Y^{(j)}(\theta_1)} & \meter{} \vcw{4} & \ket{0} \qw & \qw & \qw \\
      \lstick{$\bmc$} & \cw & \gate[cwires={1}]{\bmc(\bmh_j)} & \cwbend{2} & \cwbend{1} & \cwbend{-1} & \cwbend{-1} & \cw & \cw & \cw & \cw \\
      \lstick{$z$} & \cw & \cw & \cw & \gate[cwires={1}]{\over {\hat g}_k(\bmc_k; z)} & \cwbend{-1} & \cwbend{-1} & \cw & \cwbend{2} & \cw & \cw \\
      \lstick{$k$} & \cw & \cw & \gate[cwires={1}]{\over {\hat g}_{k,{\rm tot}}(\bmc)} & \cw & \cwbend{-1} & \cwbend{-1} & \cw & \cw & \cwbend{1} & \cw \\
      \lstick{$\bmh$} & \cwbend{-5} & \cwbend{-3} & \cw & \cw & \cw & \cw & \gate[cwires={1}]{e_j} & \gate[cwires={1}]{z_j} & \gate[cwires={1}]{k_j} & \cw
     \end{tikzcd}
  \end{adjustbox}
  \caption{
    The $j$-th step of a multi-step quantum veto simulation. The last three gates acting on the classical register $\bmh$ denote the saving of the classical information generated in the circuit into the history register.
  }
  \label{fig:veto_example_jth}
\end{figure}
}{\textbf{[Figure omitted to avoid too long compile time]}}

The veto algorithm for the multi-step simulation can be constructed very similarly to \cref{sec:gate_count}.
Consider the quantum circuit in \cref{fig:veto_example_jth}, which denotes the $j$-th step of the quantum veto simulation, and we can perform a multi-step simulation simply by connecting the same circuit repeatedly.
The major difference from \cref{fig:veto_example_fullq} is the existence of the classical history register $\bmh$ and the state and classical labels update controlled on $\bmh$.
We denote a qubit reset by the $\ket{0}$ after measurement of $\ket{e}$, and indicate the saving of the classical information generated in the step into the history register by the last three gates in the circuit.

It is straightforward to calculate the density matrix of the quantum veto simulation, and one obtains the same conclusion as \cref{sec:gate_count}, i.e., the correct density matrix is reproduced provided that the inequality \cref{eq:cond_over_fullq} is satisfied.
Furthermore, one can see that the numerical error of the approximation only accumulates linearly of the number of steps $N$, ensuring the validity of the veto procedure for the multi-step simulation.
Also, classical simulation of the multi-step quantum veto algorithm is possible as far as the state update $U(\bmh_j)$ does not introduce a complex entanglement among the state qubits.
If $n_q'$ entangled qubits are generated according to the state update, then the complexity of the classical calculations scales exponentially in $n_q'$.

\section{Parton shower and quantum Monte Carlo simulation}\label{sec:kin_QPS}

In this section, we introduce the physics model we are interested in simulating (\cref{sec:physics_model}), review the parton shower formalism (\cref{sec:parton_shower}) and describe a way to perform the Monte Carlo simulation to determine the kinematics associated with the parton shower (\cref{sec:Monte_Carlo}).
In particular, in \cref{sec:Monte_Carlo}, we show that the parton shower simulation can be regarded as an example of the general quantum Monte Carlo setup described in the previous section.

\subsection{A dark sector physics model}
\label{sec:physics_model}

The parton shower formalism is an approach to obtain the cross-section of a high-multiplicity final state where each soft and/or collinear emission is enhanced by a large logarithmic factor.
We focus on a simplified model of 2-flavor fermions with a dark $U(1)$ gauge interaction described by the Lagrangian
\begin{align}
  \mathcal{L} &=
  \bar{\chi}_1 (i\slashed{\partial} - m_{\chi}) \chi_1
  + \bar{\chi}_2 (i\slashed{\partial} - m_{\chi}) \chi_2
  - \frac{1}{4} F_{\mu\nu} F^{\mu\nu} \notag \\
  &\quad - \frac{1}{2} m_A^2 A_\mu^2
  + i\begin{pmatrix}
    \bar{\chi}_1 & \bar{\chi}_2
  \end{pmatrix} \begin{pmatrix}
    g_1 & g_{12} \\
    g_{12} & g_2
  \end{pmatrix} \slashed{A} \begin{pmatrix}
    \chi_1 \\
    \chi_2
  \end{pmatrix}\;,
  \label{eq:Lagrangian}
\end{align}
where $F_{\mu\nu} \equiv \partial_\mu A_\nu - \partial_\nu A_\mu$ is the field strength of the dark gauge boson $A_\mu$.
Note that we assume that all the fermions and gauge bosons have negligible masses compared to the energy scale of our interest, thus treating all of them as massless in the following (even though the mass is important to distinguish mass and gauge eigenbases).

In a choice of the fermion basis with $g_{12}\neq 0$, different flavors of fermions mix through the splitting processes such as $\chi_1 \to \chi_2 A$, which causes the quantum interference of diagrams of our interest.
On the other hand, for the parton shower simulation, it is useful to work in the gauge eigenbasis.
Thus, we perform a basis change with a $2 \times 2$ unitary matrix $U$ as
\begin{align}
  \begin{pmatrix}
    \chi_a \\
    \chi_b
  \end{pmatrix} = U \begin{pmatrix}
    \chi_1 \\
    \chi_2
  \end{pmatrix},
  \label{eq:basis_rotation}
\end{align}
requiring that $U$ diagonalizes the gauge coupling matrix $G$, i.e.,
\begin{align}
  G \equiv \begin{pmatrix}
    g_1 & g_{12} \\
    g_{12} & g_2
  \end{pmatrix}
  = U^\dagger \begin{pmatrix}
    \diagonalmatrix{g_a, g_b}
  \end{pmatrix} U \,,
\end{align}
with $g_a$ and $g_b$ being gauge eigenvalues.
In this basis, all emissions of bosons $A$ are flavor diagonal; we have to take account of $\chi_f \to \chi_f A$, $\bar{\chi}_f \to \bar{\chi}_f A$ and $A \to \bar{\chi}_f \chi_f$ ($f=a,b$).
Note that this basis depends on the renormalization scale since $g_1$, $g_2$, and $g_{12}$ generally have different forms of beta functions, which makes the quantum interference effect inevitable in the most general multi-flavor parton shower simulation.

\subsection{The parton shower approximation}
\label{sec:parton_shower}

The parton shower approximation uses the fact that emissions of the physics model introduced above happen predominantly at small angles or soft energies. 
It therefore expands around the small emission angle but allows for an arbitrary number of emissions, allowing it to go beyond the standard perturbative expansions.
A parton shower requires the definition of an evolution variable, which tends to zero with the emission angle. 
This evolution variable is continuously decreasing, such that an emission happening later in the shower is always guaranteed to have a value for the evolution variable that is smaller than those of any previous emissions.

We use the same parameterization of the kinematic variables as Herwig~\cite{Bahr:2008pv}, using light-cone coordinates.
Let $p_i^\mu$, $p_\ell^\mu$ and $p_m^\mu$ be the four-momenta of the partons involved in the splitting $i \to \ell m$, and define a lightlike vector $n^\mu$, which satisfies $n^0 = p_i^0$ and has space components that point towards the opposite direction to those of $p_i^\mu$.
This allows to decompose the final state momenta as
\begin{align}
  p_\ell &= \eta_\ell p_i + \zeta_\ell n + p_{\perp \ell}\,, \\
  p_m &= \eta_m p_i + \zeta_m n + p_{\perp m}\,, 
\end{align}
where $p_{\perp \ell}$ and $p_{\perp m}$ are four-momenta orthogonal to both $p_i$ and $n$.
Using these variables, we define the energy fraction $z$ and the virtuality $\qSq$ of the splitting $i\to \ell m$ as
\begin{align}
  z &\equiv \frac{\eta_\ell}{\eta_\ell + \eta_m}\,, \\
  \qSq &\equiv - \frac{\left( p_\ell - z (p_\ell + p_m) \right)^2}{z^2 (1-z)^2}\,.
\end{align}

The probability densities of the splitting processes are given as functions of the virtuality $\qSq$ by
\begin{align}
  R_{ff}(\qSq) &= \frac{1}{\qSq} \int_{z_{-}(\qSq)}^{z_{+}(\qSq)} dz\,
  \frac{\alpha_f(\mu_{\chi\chi}(\qSq, z))}{2\pi}
  P_{\chi\chi}(z)\,,
  \notag \\
  R_{f A}(\qSq) &= \frac{1}{\qSq} \int_{z_{-}(\qSq)}^{z_{+}(\qSq)} dz\,
  \frac{\alpha_f(\mu_{\chi A}(\qSq, z))}{2\pi}
  P_{\chi A}(z)\,,
\end{align}
where $ R_{ff}$ and $R_{fA}$ respectively correspond to the splitting processes $\chi_f \to \chi_f A$ (or $\bar{\chi}_f \to \bar{\chi}_f A$) and $A\to \bar{\chi}_f \chi_f$ with $f=a,b$.
$z_\pm (\qSq)$ are the minimum and maximum of the energy fraction allowed by the kinematics for a fixed value of $\qSq$, $\alpha_f = g_f^2/4\pi$, and $\mu_{\chi\chi}$ and $\mu_{\chi A}$ are the renormalization scales.

It has been shown~\cite{Bassetto:1983mvz, Amati:1980ch} that the proper choices of the renormalization scale are
\begin{align}
  \mu_{\chi\chi}^2 (\qSq, z) = z (1-z)^2 \qSq + \Lambda_D^2\,, \\
  \mu_{\chi A}^2 (\qSq, z) = z^2 (1-z)^2 \qSq + \Lambda_D^2\,,
\end{align}
where $\Lambda_D \sim m_\chi \sim m_A$ is the typical mass scale of the dark sector Lagrangian.
In the above expressions, the last terms $\Lambda_D^2$ are prescriptions similar to \cite{Bierlich:2022pfr} that ensures the renormalization scale to be larger than the IR cutoff scale.
The Altarelli-Parisi splitting functions~\cite{ALTARELLI1977298} are given by
\begin{align}
  P_{\chi\chi}(z) &= \frac{1+z^2}{1-z}\,, \\
  P_{\chi A}(z) &= z^2 + (1-z)^2\,.
\end{align}
Finally, the Sudakov factor, which represents the probability at which a parton does not undergo a splitting within $\qSq \in [\qSq_1, \qSq_2]$, is obtained as
\begin{align}
  \Delta_f (\qSq_1; \qSq_2) &= \exp \left[
    - \int_{\qSq_1}^{\qSq_2} d\qSq\, R_{ff}(\qSq)
  \right]\,, \\
  \Delta_A (\qSq_1; \qSq_2) &= \exp \left[
    - \int_{\qSq_1}^{\qSq_2} d\qSq\, \left(
      R_{a A}(\qSq) + R_{b A}(\qSq)
    \right)
  \right]\,.
\end{align}

When an initial parton is generated at the energy scale $E_0$, a splitting of the parton should have the virtuality $\qSq \lesssim E_0^2$.
On the other hand, the minimum virtuality is roughly given by the dark sector scale $\Lambda_D^2$.
Note that, since each splitting $i\to \ell m$ with $(\qSq, z)$ splits the energy of the initial parton $i$ into two final partons $\ell$ and $m$, the maximum virtuality of any further splittings becomes $z^2 \qSq$ ($(1-z)^2 \qSq$) for the parton $\ell$ ($m$).
Thus, the Monte Carlo simulation of the parton shower, which will be outlined below, is a procedure to determine sets of kinematic variables $(\qSq, z)$ for splittings, scanning the associated virtuality variables from $E_0^2$ to $\Lambda_D^2$.

\subsection{Monte Carlo simulation}
\label{sec:Monte_Carlo}

For the Monte Carlo simulation of the parton shower on quantum computers, we use the results of~\cref{sec:Toy_model} to construct a vetoed version of the QPS algorithm originally introduced in \cite{Bauer:2019qxa} and improved in \cite{Deliyannis:2022uyh}.
In this approach, one discretizes the virtuality variables 
\begin{align}
    \qSq \to \qSq_j\,, \qquad j = 0 \ldots N-1
    \,,
\end{align}
with a constant step ratio 
\begin{align}
    r = \frac{\qSq_{j+1}}{\qSq_j} < 1
    \,.
\end{align}
One then judges if a splitting occurs at each step labelled by $j$.
In this work, we assume that the virtuality range defined by the value of $r$ is narrow enough such that the multi-emission probability within one step is negligible.
Defining $k=1,\dots,n$ as a label of partons with $n$ the total number, we define the virtuality of the $k$'th parton as $\qSq_j(k)$ and the allowed range at the $j$'th step is $\qSq \in \left[ \qSqj(k), r \, \qSqj(k) \right]$.

The total emission probability in the $j$-th step depends on the particle flavors $S_k = a, b, A$, as well as on the values $\qSqj(k)$ of the step $j$.
We can write
\begin{align}
  f^j_{\rm tot}(\{S_k\}, \{\qSqj(k)\}) \simeq \sum_{k=1}^n g^j_{k, {\rm tot}}(S_k, \qSqj(k)) 
  \,,
\end{align}
where we have defined
\begin{align}
    g^j
    _{k, {\rm tot}}(S_k, \qSqj(k)) = \left(
    1 - \Delta_{S_k} (\qSqj(k) ; r \, \qSqj(k))
  \right)\,.
\end{align}
As long as the step size is chosen to be small enough one can expand the exponential, and evaluate the virtuality integral analytically, resulting in the emission probability
\begin{align}
  g^j_{k, {\rm tot}}(S_k, \qSqj(k)) \simeq& \int_{z_- \left( \qSq_j(k) \right)}^{z_+ \left( \qSq_j(k) \right)} \!\!\! dz \, g^j_k(S_k, \qSqj(k); z)
  \label{eq:P_analogous_1}
  \,,
\end{align}
with
\begin{align}
  g&^j_{k}(S_k, \qSqj(k); z) \simeq -\ln r \notag\\
  & \bigg[
    \left. 
      \frac{(\alpha_a (\mu) \delta_{S_k, a}
      + \alpha_b (\mu) \delta_{S_k, b})}{2\pi}
    \right|_{\mu = \mu_{\chi\chi} \left( \qSqj(k), z \right)}
    P_{\chi\chi}(z) \notag \\
    & \,\, +\left.
      \delta_{S_k, A}\,\frac{(\alpha_a (\mu) + \alpha_b (\mu))}{2\pi}
    \right|_{\mu = \mu_{\chi A} \left( \qSqj(k), z \right)}
    P_{\chi A}(z) 
  \bigg]\,.
  \label{eq:P_analogous_2}
\end{align}

Given these expressions, one can see that there are two types of emission probabilities. 
The first is the probability for a fermion labeled by $k$ to radiate a boson.
This emission depends on the flavor of the fermion ($a$ or $b$), which we keep track of with a quantum variable $q_k$ in order to be able to take into account the interference between these two flavors. 
The emission probability also depends on the value of $\qSqj(k)$, which we keep track of with a classical variable. 
The emission probability is given by
\begin{align}
    g_{k,\chi}^j\left(q_k, \qSqj(k); z\right) = (-\ln r) \frac{\alpha_{q_k}(\mu_{\chi\chi})}{2\pi}\, P_{\chi\chi}(z)
    \,,
\end{align}
where $\mu_{\chi\chi}$ is shorthand for $\mu_{\chi\chi}(\qSqj(k), z)$. 
The second is the probability of a boson to split into two fermions. Since there is only a single boson present in the dark sector physics model, this splitting probability does not depend on a quantum variable, and only depends on the classical information $\qSqj(k)$
\begin{align}
    g_{k,A}^j\left(\qSqj(k); z\right) = (-\ln r)\frac{\alpha_{a}(\mu_{\chi A})+\alpha_{b}(\mu_{\chi A})}{2\pi}\, P_{\chi A}(z) 
    \,,
\end{align}
where $\mu_{\chi A}$ is shorthand for $\mu_{\chi A}(\qSqj(k), z)$. 
Note that in the splitting $A \to \chi\chi$ one needs to introduce a quantum correlation between the fermions generated, and this will be discussed in more detail later.

From the discussion so far, we see that the problem has a very similar form as the toy model we have introduced in~\cref{sec:Toy_model}, with the only difference being that one has to differentiate if the $k$'th particle denotes a fermion $\chi$ or a boson $A$. 
Given that the value $k$ is determined in the QVPS algorithm classically, and the type of particle $t_k=\chi, A$ can be kept track of with classical information, this only adds an additional step to the classical processing required in the algorithm.

For each $k$ we now introduce overestimated quantities with the following conditions:\footnote{
  Note that $\zmover$, $\zpover$, and $\alpover$ could be defined as functions of $\qSq$.
  In this paper, we use constant values instead to be consistent with the treatment in \cite{Bahr:2008pv}.
}
\begin{itemize}
  \item $\left[ \zmover, \zpover \right] \supseteq
  \left[ z_-(\qSq), z_+(\qSq) \right]$ $\quad (\forall \qSq)$\,,
  \item $\alpover > \max \left( \alpha_f \left( \mu_{\chi\chi} (\qSq, z) \right), \alpha_f \left( \mu_{\chi A} (\qSq, z) \right) \right)$
  \\ $\quad (\forall \qSq$, $z$, $f)$\,,
  \item $\Pover[\chi] (z) > P_{\chi\chi} (z)$ $\quad (\forall z)$\,,
  \item $\Pover[A] (z) > P_{\chi A}(z)$ $\quad (\forall z)$\,.
\end{itemize}
The overestimated probability density is independent of the flavor of the fermion, which implies that the overestimated probability density can be written as a function of classical information $\bmc_k = \left\{ t_k, \qSqj(k) \right\}$ as
\begin{align}
    g_k^{j, {\rm over}} \left(
      \bmc_k ; z \right) & =  g_{k,\chi}^{j, {\rm over}} \left(
    \qSq_j(k) ; z \right) \delta_{t_k, \chi}\notag\\
    &\quad + g_{k,A}^{j, {\rm over}} \left(\qSq_j(k); z \right) \delta_{t_k, A}
    \,.
\end{align}
The functions $g_{k,{t_k}}^{j, {\rm over}}$ have the simple expression
\begin{align}
  g_{k,\chi}^{j, {\rm over}} \left(
    \qSq_j(k) ; z
  \right) &= (-\ln r) \frac{\over{\alpha}}{2\pi} \over{P}_\chi (z)
  \,,
  \notag\\
    g_{k,A}^{j, {\rm over}} \left(
    \qSq_j(k) ; z
  \right) &= (-\ln r) \frac{\over{\alpha}}{\pi} \over{P}_A (z)
  \,.
\end{align}
We also define
\begin{align}
    g_{k, {\rm tot}}^{j, {\rm over}} \left(
    \bmc_k \right) \equiv \int_{z_-^{\rm over}}^{z_+^{\rm over}}\! {\rm d} z \, g_k^{j, {\rm over}} \left(
    \bmc_k ; z \right)
    \,,
\end{align}
as well as
\begin{align}
    \hat g_k^{j, {\rm over}} \left(\bmc_k ; z \right) \equiv \frac{g_k^{j, {\rm over}} \left(\bmc_k ; z \right)}{g_{k, {\rm tot}}^{j, {\rm over}} \left(
      \bmc_k \right)}
    \,,
\end{align}
and
\begin{align}
    \hat g_{k, {\rm tot}}^{j, {\rm over}} \left(
      \bmc \right) \equiv \frac{g_{k, {\rm tot}}^{j, {\rm over}} \left(
        \bmc_k \right)}{\sum_k g_{k, {\rm tot}}^{j, {\rm over}} \left(
          \bmc_k\right)}
    \,.
\end{align}

\subsection{The QVPS algorithm}
\label{sec:QVPS_algorithm}
In this section, we will describe the QVPS algorithm that can be used to shower a set of particles according to the physics model of~\cref{sec:physics_model} and include the full quantum interference from the different flavors of fermions.
For the model discussed in this paper, there are two fermion flavors possible, which can be captured by one qubit for each fermion. 
The basic layout of the algorithm has a very similar general structure as the original QPS algorithm by of~\cite{Bauer:2019qxa,Deliyannis:2022uyh}.
It is divided into $N$ steps representing discretized values of the evolution variable, where at each step either an emission occurs or it does not.
If the parton shower algorithm has $n_I$ partons present initially, it can contain up to $n_{\rm max} = n_I + N$ partons at the end of the evolution, since in principle each step can produce one additional parton due to the $\chi \to \chi A$ or the $A \to \chi \chi$ splittings.  
At the beginning of each step, the quantum register holding the fermion information is rotated from the mass eigenstates to the interaction eigenstates.
This rotation depends on the running coupling constants, which in turn depend on the kinematics of the previous emissions, and is therefore different at each step.
After this rotation, one uses the QVPS algorithm to decide if an emission occurred at the given discrete value of the evolution variable and if it did, find the correct $z$ value of the emission. 

Since in the QVPS algorithm the value of $k$, which determines the potential particle to radiate, is determined classically through the overestimated probability distributions, each step in the QVPS algorithm depends only on the quantum state of one of the $n_{\rm max}$ qubits holding the fermion flavor information. 
For each step $j$, the corresponding quantum algorithms is given by the following circuit.
\ifthenelse{\boolean{compileAll}}{
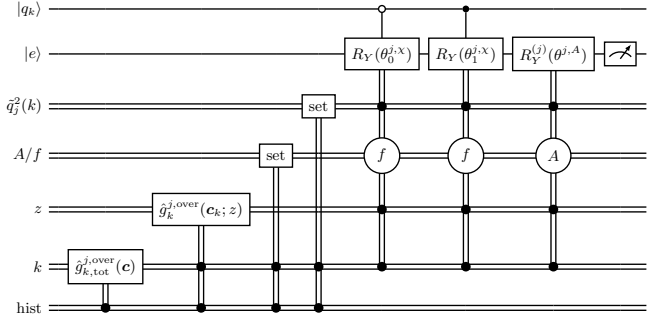
\begin{figure}[htp]
  \centering
  \begin{adjustbox}{width=\hsize}
    \begin{tikzcd}[column sep=0.2cm]
      \lstick{$\ket{q_k}$} & \qw & \qw & \qw & \qw & \qw & \octrl{1} & \ctrl{1} & \qw & \qw & \qw \\
      \lstick{$\ket{e}$} & \qw & \qw & \qw & \qw & \qw & \gate{R_Y(\theta_0^{j,\chi})} &  \gate{R_Y(\theta_1^{j,\chi})} & \gate{R_Y^{(j)}(\theta^{j,A})} & \meter{} & \qw\\
      \lstick{$\qSqj(k)$} & \cw & \cw & \cw & \cw & \gate[cwires={1}]{\rm set} & \cwbend{-1} & \cwbend{-1} & \cwbend{-1} & \cw & \cw\\
      \lstick{$A/f$} & \cw & \cw & \cw & \gate[cwires={1}]{\rm set} & \cw & \gate[cwires={1},style={circle}]{f} & \gate[cwires={1},style={circle}]{f} & \gate[cwires={1},style={circle}]{A} & \cw & \cw\\
      \lstick{$z$} & \cw & \cw & \gate[cwires={1}]{{\hat g}_k^{j, {\rm over}}(\bmc_k; z)} & \cw & \cw &  \cwbend{-2} & \cwbend{-2} & \cwbend{-2} & \cw & \cw\\
      \lstick{$k$} & \cw & \gate[cwires={1}]{{\hat g}_{k,{\rm tot}}^{j, {\rm over}}(\bmc)} & \cwbend{-1} & \cwbend{-2} & \cwbend{-3} &  \cwbend{-1} & \cwbend{-1} & \cwbend{-1} & \cw & \cw\\
      \lstick{${\rm hist}$} & \cw & \cwbend{-1} & \cwbend{-1} & \cwbend{-1} & \cwbend{-1} & \cw & \cw & \cw & \cw & \cw
     \end{tikzcd}
  \end{adjustbox}
  \caption{
    A single step of a quantum veto algorithm according to~\cref{eq:splitting_op} that determines the kinematics of a possible emission. Note that each step will in general have a different classical value $k$, and will therefore be controlled on a different qubit $\ket{q_k}$. 
  }
\end{figure}
}{\textbf{[Figure omitted to avoid too long compile time]}}
It starts with a range of classical calculations. The first is selecting a value of $k$ according to the probability ${\hat g}_{k,{\rm tot}}^{j, {\rm over}}(\bmc) = \hat g_{k,{\rm tot}}^{j, {\rm over}} \left( \left\{ t_k, \qSqj(k) \right\}_k \right)$. After the value of $k$ has been determined, one selects a value of $z$ based on the probability distribution $\hat g_k^{j, {\rm over}} \left( \bmc_k ; z \right) = \hat g_k^{j, {\rm over}} \left(t_k, \qSq_j(k) ; z \right)$.
Note that this of course requires for the overestimated distributions to be chosen to be simple enough to make this sampling possible.
Finally, using the information of the history one determines if the emitting particle $k$ is a fermion $\chi$ or boson $A$, and the value of $\qSqj(k)$ to be used for the veto procedure. 

After this classical processing, one applies rotations to the emission qubit $\ket{e}$, which correct from the overestimated probabilities to probability distribution $g^j_{k}(S_k, \qSqj(k); z)$, which depend on the flavor of the emitting particle. 
For the case of the emitting particle being a fermion, the rotation is controlled on the qubit $\ket{q_k}$ holding the fermion flavor information. The rotation angle also depends on the values of $z$ and $\qSqj(k)$ stored in classical registers. 
The rotation angles are given by
\begin{align}
\begin{split}
    \theta_0^{j,\chi} & = \hat \Theta \left( \frac{g^j
    _k(a, \qSqj(k) ; z)}{\hat g_{k, {\rm tot}}^{j, {\rm over}} \left(
    \bmc\right) \hat{g}_k^{j, {\rm over}} \left(
    \chi, \qSq_j(k) ; z\right)}\right) \,,
    \\
    \theta_1^{j,\chi} & = \hat \Theta \left( \frac{g^j
    _k(b, \qSqj(k) ; z)}{\hat g_{k, {\rm tot}}^{j, {\rm over}} \left(
    \bmc\right) \hat{g}_k^{j, {\rm over}} \left(
    \chi, \qSq_j(k) ; z\right)}\right) \,,
    \\
    \theta^{j,A} & = \hat \Theta \left( \frac{g^j
    _k(A, \qSqj(k) ; z)}{\hat g_{k, {\rm tot}}^{j, {\rm over}} \left(
    \bmc\right) \hat{g}_k^{j, {\rm over}} \left(
    A, \qSq_j(k) ; z\right)}\right) \,.
\end{split}
\end{align}

After these rotations are performed, one measures the qubit $\ket{e}$ and updates the history register with the information of whether a splitting has occurred, and if yes, which particle emitted and with what emission kinematics.
If an emission occurred from a gauge boson $A$ (implying that $k$ corresponded to a boson $A$ and $\ket{e}$ was measured to be $\ket{1}$), an additional step is required since one needs to add two extra fermions to the quantum register.
Assuming we had $n$ fermions before the emission from a boson in a given step, we need to add two fermions to the event, which are entangled according to the coupling constants of fermions $a$ and $b$.
This is accomplished through the additional circuit in \cref{fig:additional},
where the rotation angle $\Phi$ is given by
\begin{align}
    \Phi= 2 \arccos\left( \frac{\alpha_a(\mu)}{\sqrt{\alpha_a^2(\mu) + \alpha_b^2(\mu)}}\right)
    \,.
\end{align}

\ifthenelse{\boolean{compileAll}}{
\begin{figure}[tp!]
  \centering
  \begin{adjustbox}{height=2cm}
    \begin{tikzcd}[column sep=0.2cm]
      \lstick{$\ket{q_{n+2}}$} & \qw & \gate{R(\Phi)} & \ctrl{1} & \qw\\
      \lstick{$\ket{q_{n+1}}$} & \qw & \qw & \targ{} & \qw\\
      \lstick{$\ket{e}$} & \qw & \qw & \qw & \qw\\
      \lstick{$\qSqj(k)$} & \cw & \cwbend{-3} & \cw & \cw\\
      \lstick{$A/f$} & \cw & \gate[cwires={1},style={circle}]{A} & \cw & \cw \\
      \lstick{$z$} & \cw & \cwbend{-2} & \cw & \cw\\
      \lstick{$k$} & \cw & \cw & \cw & \cw \\
      \lstick{${\rm hist}$} & \cw  & \cw & \cw & \cw
     \end{tikzcd}
  \end{adjustbox}
  \caption{
    The quantum circuit that adds two extra fermions to the event with the correct entanglement. The angle $\Phi$ is defined in the text.
  }
  \label{fig:additional}
\end{figure}
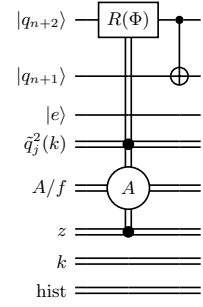
}{\textbf{[Figure omitted to avoid too long compile time]}}

\subsection{Numerical results}
\label{sec:numerical_results}
In order to illustrate the potential of the QVPS algorithm to simulate the quantum mechanical effects on parton shower dynamics, we perform a parton shower simulation with a one-fermion initial state $\chi_1$.
We use the initial value of virtuality $\qSq_0 = 100$ and the dark scale $\Lambda_D = 0.1$.
We further assume that the rotation matrix $U$ in \cref{eq:basis_rotation} is expressed scale-independently as
\begin{align}
  U = \frac{1}{\sqrt{2}} \begin{pmatrix}
    1 & 1 \\
    1 & -1
  \end{pmatrix}
  \;,
\end{align}
while the gauge eigenvalues run according to
\begin{align}
  \alpha_f(\mu) &= \frac{\alpha_f^0}{\beta_0 \ln (\mu^2/\Lambda_D^2)}
  \;,\\
  \beta_0 &= \frac{33-2 n_f}{4\pi}
  \;,
\end{align}
with $\alpha_f^0 = 0.5, 2$ for $f=a, b$, respectively, while $n_f=2$.
Although we start with the $\chi_1$ state, the quantum mechanical effect during the shower process rotates the state, and there is a non-zero probability of finding the $\chi_2$ state of the corresponding fermion in the end.
This probability should depend on the number of emissions.

\begin{figure}[t]
  \centering
  \includegraphics[width=\hsize]{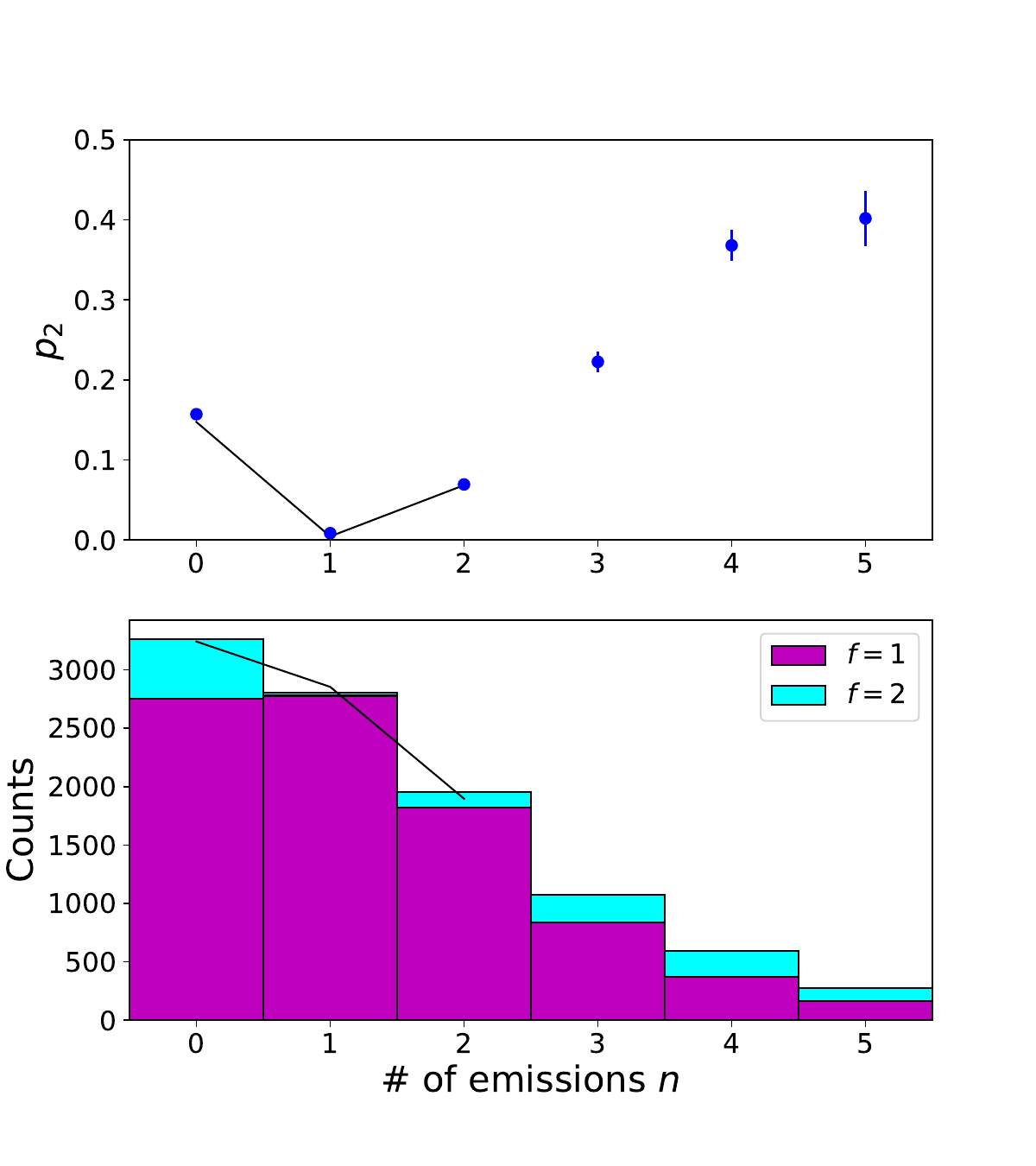}
  \caption{
    Final state fermion flavor information as a function of the number of emissions. 
    Two panels show the probability of finding the $\chi_2$ state after showering (top) and the stacked histogram of events with $\chi_1$ and $\chi_2$ final states (bottom).
    Blue data points and histograms show the numerical results, while the black solid lines show the analytic ones.
  }
  \label{fig:flavor_change}
\end{figure}

In \cref{fig:flavor_change}, we show the numerical results with the total number of shots $N_{\mathrm{shots}} = 10^4$.
In the top panel, we plot with blue dots the probability $p_2$ at which we find the $\chi_2$ final state after the observation as a function of the number of emissions $n$.
The error bars correspond to the statistical errors.
In the bottom panel, we show the stacked histogram of events with the $\chi_1$ and $\chi_2$ final states with the purple and cyan colors, respectively.
We can see non-zero and $n$-dependent $p_2$ as expected.

By explicitly performing the numerical integration, we can obtain the analytic estimation of these values.
Firstly, the distribution of $n$ is obtained with using the Sudakov factors $\Delta_f$ and the splitting probability densities $R_{ff}(\qSq)$ with $f=a,b$.
For example, for $n=0$ and $1$, we obtain the expected number of counts in each bin as $N(n) = \frac{1}{2} (N_a(n) + N_b(n))$, where
\begin{align}
  N_f(n=0) &= N_{\mathrm{shots}} \Delta_f \left( \Lambda_D^2 ; \qSq_0 \right)
  \;, \\
  N_f(n=1) &= N_{\mathrm{shots}} \int_{\Lambda_D^2}^{\qSq_0} d\qSq\, \int_{z_-(\qSq)}^{z_+(\qSq)} dz\,
  \Delta_f \left( \Lambda_D^2 ; z^2 \qSq \right) \nonumber\\
  & \quad \times \Delta_A \left( \Lambda_D^2 ; (1-z)^2 \qSq \right)
  \frac{1}{\qSq} \nonumber\\
  & \quad \times \frac{\alpha_f(\mu_{\chi\chi}(\qSq, z))}{2\pi} P_{\chi\chi}(z) \Delta_f \left( \qSq ; \qSq_0 \right)
  \;,
\end{align}
with $f=a,b$.
Note that we define $\Delta_{f/A}(\qSq_1 ; \qSq_2)=1$ when $\qSq_1 > \qSq_2$.
Using them, the probability $p_2$ can also be calculated as a function of $n$ as
\begin{align}
  p_2(n) = \frac{\left( \sqrt{N_a(n)} - \sqrt{N_b(n)} \right)^2}{2(N_a(n) + N_b(n))},
\end{align}
where the correlation term in the numerator clearly exhibits the interference effect between two fermion flavors.
In \cref{fig:flavor_change}, the analytic results up to $n\leq 2$ are shown in black solid lines.
It can be seen that the numerical results from the QVPS algorithm agree with the analytic ones well within the statistical errors.
Note that the computational resource required for the analytic estimation of $N_f$ and $p_2$ grows exponentially as $n$ increases since we have to evaluate two additional integrals for each additional emission.
On the other hand, the computational resource required for the QVPS algorithm only grows linearly in $n$, which clearly demonstrates its potential for processes with many emissions.

%
\section{Conclusions} \label{sec:conclusions}
%

In this paper we introduced the quantum veto algorithm (QVPS) for 
the quantum parton shower. This is a genuinely quantum-classical hybrid algorithm, and its efficiency is significantly improved 
from previous quantum algorithms in the literature. In particular, for initial states with low entanglement, we can even efficiently simulate the quantum parton showers classically. 

We expect that our quantum-classical hybrid implementation of the veto procedure will have wider applicability whenever we sample from probability distributions determined by quantum states. It would be interesting to apply our quantum veto algorithm to many more problems in different contexts.

\begin{acknowledgments}

Quantum circuit diagrams in this paper are generated using the public code \texttt{Quantikz}~\cite{Kay:2018huf}.
CWB and SC were supported by the DOE, Office of Science under contract DE-AC02-05CH11231, partially through Quantum Information Science Enabled Discovery (QuantISED) for High Energy Physics (KA2401032). MY was supported in part by the JSPS Grant-in-Aid for Scientific Research (19H00689, 19K03820, 20H05860, 23H01168), and by JST, Japan (PRESTO Grant No.\ JPMJPR225A, Moonshot R\&D Grant No.\ JPMJMS2061).
CWB and MY also acknowledge support by the Munich Institute for Astro-, Particle and BioPhysics (MIAPbP) which is funded by the Deutsche Forschungsgemeinschaft (DFG, German Research Foundation) under Germany's Excellence Strategy-EXC-2094-390783311.
\end{acknowledgments}

\bibliography{bib}

\providecommand{\href}[2]{#2}\begingroup\raggedright\begin{thebibliography}{10}

\bibitem{Bauer:2022hpo}
C.~W. Bauer {\em et~al.}, ``{Quantum Simulation for High-Energy Physics},''
  \href{http://dx.doi.org/10.1103/PRXQuantum.4.027001}{{\em PRX Quantum}
  {\bfseries 4} no.~2, (2023) 027001},
  \href{http://arxiv.org/abs/2204.03381}{{\ttfamily arXiv:2204.03381
  [quant-ph]}}.

\bibitem{nielsen2010quantum}
M.~A. Nielsen and I.~L. Chuang, {\em Quantum computation and quantum
  information}.
\newblock Cambridge university press, 2010.

\bibitem{Bauer:2021gup}
C.~W. Bauer, M.~Freytsis, and B.~Nachman, ``{Simulating Collider Physics on
  Quantum Computers Using Effective Field Theories},''
  \href{http://dx.doi.org/10.1103/PhysRevLett.127.212001}{{\em Phys. Rev.
  Lett.} {\bfseries 127} no.~21, (2021) 212001},
  \href{http://arxiv.org/abs/2102.05044}{{\ttfamily arXiv:2102.05044
  [hep-ph]}}.

\bibitem{Collins:1987pm}
J.~C. Collins and D.~E. Soper, ``{The Theorems of Perturbative QCD},''
  \href{http://dx.doi.org/10.1146/annurev.ns.37.120187.002123}{{\em Ann. Rev.
  Nucl. Part. Sci.} {\bfseries 37} (1987) 383--409}.

\bibitem{Sterman:1995fz}
G.~F. Sterman, ``{Partons, factorization and resummation, TASI 95},'' in {\em
  Theoretical Advanced Study Institute in Elementary Particle Physics (TASI
  95): QCD and Beyond}, pp.~327--408.
\newblock 6, 1995.
\newblock \href{http://arxiv.org/abs/hep-ph/9606312}{{\ttfamily
  arXiv:hep-ph/9606312}}.

\bibitem{Jaffe:1996zw}
R.~L. Jaffe, ``{Spin, twist and hadron structure in deep inelastic
  processes},'' in {\em Ettore Majorana International School of Nucleon
  Structure: 1st Course: The Spin Structure of the Nucleon}, pp.~42--129.
\newblock 1, 1996.
\newblock \href{http://arxiv.org/abs/hep-ph/9602236}{{\ttfamily
  arXiv:hep-ph/9602236}}.

\bibitem{Bauer:2002nz}
C.~W. Bauer, S.~Fleming, D.~Pirjol, I.~Z. Rothstein, and I.~W. Stewart, ``{Hard
  scattering factorization from effective field theory},''
  \href{http://dx.doi.org/10.1103/PhysRevD.66.014017}{{\em Phys. Rev. D}
  {\bfseries 66} (2002) 014017},
  \href{http://arxiv.org/abs/hep-ph/0202088}{{\ttfamily arXiv:hep-ph/0202088}}.

\bibitem{Bauer:2019qxa}
C.~W. Bauer, W.~A. de~Jong, B.~Nachman, and D.~Provasoli, ``{Quantum Algorithm
  for High Energy Physics Simulations},''
  \href{http://dx.doi.org/10.1103/PhysRevLett.126.062001}{{\em Phys. Rev.
  Lett.} {\bfseries 126} no.~6, (2021) 062001},
  \href{http://arxiv.org/abs/1904.03196}{{\ttfamily arXiv:1904.03196
  [hep-ph]}}.

\bibitem{Bepari:2020xqi}
K.~Bepari, S.~Malik, M.~Spannowsky, and S.~Williams, ``{Towards a quantum
  computing algorithm for helicity amplitudes and parton showers},''
  \href{http://dx.doi.org/10.1103/PhysRevD.103.076020}{{\em Phys. Rev. D}
  {\bfseries 103} no.~7, (2021) 076020},
  \href{http://arxiv.org/abs/2010.00046}{{\ttfamily arXiv:2010.00046
  [hep-ph]}}.

\bibitem{Li:2021kcs}
{\bfseries QuNu} Collaboration, T.~Li, X.~Guo, W.~K. Lai, X.~Liu, E.~Wang,
  H.~Xing, D.-B. Zhang, and S.-L. Zhu, ``{Partonic collinear structure by
  quantum computing},''
  \href{http://dx.doi.org/10.1103/PhysRevD.105.L111502}{{\em Phys. Rev. D}
  {\bfseries 105} no.~11, (2022) L111502},
  \href{http://arxiv.org/abs/2106.03865}{{\ttfamily arXiv:2106.03865
  [hep-ph]}}.

\bibitem{Bepari:2021kwv}
K.~Bepari, S.~Malik, M.~Spannowsky, and S.~Williams, ``{Quantum walk approach
  to simulating parton showers},''
  \href{http://dx.doi.org/10.1103/PhysRevD.106.056002}{{\em Phys. Rev. D}
  {\bfseries 106} no.~5, (2022) 056002},
  \href{http://arxiv.org/abs/2109.13975}{{\ttfamily arXiv:2109.13975
  [hep-ph]}}.

\bibitem{Macaluso:2021ngq}
S.~Macaluso and K.~Cranmer, ``{The Quantum Trellis: A classical algorithm for
  sampling the parton shower with interference effects},''
  \href{http://arxiv.org/abs/2112.12795}{{\ttfamily arXiv:2112.12795
  [hep-ph]}}.

\bibitem{Chigusa:2022act}
S.~Chigusa and M.~Yamazaki, ``{Quantum simulations of dark sector showers},''
  \href{http://dx.doi.org/10.1016/j.physletb.2022.137466}{{\em Phys. Lett. B}
  {\bfseries 834} (2022) 137466},
  \href{http://arxiv.org/abs/2204.12500}{{\ttfamily arXiv:2204.12500
  [hep-ph]}}.

\bibitem{Deliyannis:2022uyh}
P.~Deliyannis, J.~Sud, D.~Chamaki, Z.~Webb-Mack, C.~W. Bauer, and B.~Nachman,
  ``{Improving quantum simulation efficiency of final state radiation with
  dynamic quantum circuits},''
  \href{http://dx.doi.org/10.1103/PhysRevD.106.036007}{{\em Phys. Rev. D}
  {\bfseries 106} no.~3, (2022) 036007},
  \href{http://arxiv.org/abs/2203.10018}{{\ttfamily arXiv:2203.10018
  [hep-ph]}}.

\bibitem{Bahr:2008pv}
M.~Bahr {\em et~al.}, ``{Herwig++ Physics and Manual},''
  \href{http://dx.doi.org/10.1140/epjc/s10052-008-0798-9}{{\em Eur. Phys. J. C}
  {\bfseries 58} (2008) 639--707},
  \href{http://arxiv.org/abs/0803.0883}{{\ttfamily arXiv:0803.0883 [hep-ph]}}.

\bibitem{Bassetto:1983mvz}
A.~Bassetto, M.~Ciafaloni, and G.~Marchesini, ``{Jet Structure and Infrared
  Sensitive Quantities in Perturbative QCD},''
  \href{http://dx.doi.org/10.1016/0370-1573(83)90083-2}{{\em Phys. Rept.}
  {\bfseries 100} (1983) 201--272}.

\bibitem{Amati:1980ch}
D.~Amati, A.~Bassetto, M.~Ciafaloni, G.~Marchesini, and G.~Veneziano, ``{A
  Treatment of Hard Processes Sensitive to the Infrared Structure of QCD},''
  \href{http://dx.doi.org/10.1016/0550-3213(80)90012-7}{{\em Nucl. Phys. B}
  {\bfseries 173} (1980) 429--455}.

\bibitem{Bierlich:2022pfr}
C.~Bierlich {\em et~al.}, ``{A comprehensive guide to the physics and usage of
  PYTHIA 8.3}'' \href{http://arxiv.org/abs/2203.11601}{{\ttfamily
  arXiv:2203.11601 [hep-ph]}}.

\bibitem{ALTARELLI1977298}
G.~Altarelli and G.~Parisi, ``Asymptotic freedom in parton language,''
  \href{http://dx.doi.org/https://doi.org/10.1016/0550-3213(77)90384-4}{{\em
  Nuclear Physics B} {\bfseries 126} no.~2, (1977) 298--318}.
  \url{https://www.sciencedirect.com/science/article/pii/0550321377903844}.

\bibitem{Kay:2018huf}
A.~Kay, ``{Tutorial on the Quantikz Package},''
  \href{http://arxiv.org/abs/1809.03842}{{\ttfamily arXiv:1809.03842
  [quant-ph]}}.

\end{thebibliography}\endgroup
\bibliographystyle{utphys}

\end{document}